\documentclass[twocolumn,floatfix,showpacs,tightenlines,superscriptaddress]{revtex4-2}
\usepackage{mathtools}
\usepackage{bm}
\usepackage{dsfont,amsthm,amsbsy}
\usepackage{verbatim}
\usepackage{amssymb}
\usepackage{amsmath}
\usepackage{bbm}
\usepackage{graphicx}
\usepackage{epstopdf}
\usepackage{subfigure}
\usepackage{natbib}
\usepackage{epsfig}
\usepackage{amsfonts}
\usepackage{mathrsfs}
\usepackage{sidecap}
\usepackage{lipsum}
\usepackage[toc,page,title,titletoc,header]{appendix}
\usepackage[colorlinks,linkcolor=blue,citecolor=blue,anchorcolor=blue,urlcolor=blue]{hyperref}
\usepackage{hyperref}
\usepackage{resizegather}
\usepackage{tikz}
\usepackage{float}
\usepackage{mathbbol}
\usepackage[normalem]{ulem}
\usepackage{cancel}
\usepackage{upgreek}

\usepackage{dblfloatfix}
\usepackage{graphicx}  
\usepackage{dcolumn}   
\usepackage{bm}        
\usepackage{amssymb}   
\usepackage{amsmath}
\usepackage{lipsum}
\usepackage{xcolor}
\usepackage{ulem}
\usepackage{braket}

\begin{document}

\title{Emergence of Topological Non-Fermi Liquid Phases in a Modified Su–Schrieffer–Heeger Chain with Long-Range Interactions}
\author{Sepide Mohamadi}
\affiliation{Department of Physics, Institute for Advanced Studies in Basics Sciences (IASBS), Zanjan 45137- 66731, Iran}
\author{Jahanfar Abouie}
\email{jahan@iasbs.ac.ir}
\affiliation{Department of Physics, Institute for Advanced Studies in Basics Sciences (IASBS), Zanjan 45137- 66731, Iran}

\date{\today}

\begin{abstract}
In this study, we investigate the emergence of a topological non-Fermi liquid (NFL) phase in a modified Su–Schrieffer–Heeger (SSH) chain model subjected to long-range interactions characterized by the Hatsugai-Kohmoto (HK) model. While Fermi liquid theory has been instrumental in understanding low-temperature properties of metals, it fails to account for the complex behaviors exhibited by strongly correlated systems, where interactions lead to emergent phenomena such as non-Fermi liquid behavior. Our analysis reveals that the SSH-HK model supports a rich ground state phase diagram, exhibiting distinct NFL phases marked by many-body Zak phases of $2\pi$ and $0$, corresponding to topological and trivial NFL states, respectively. We demonstrate that the topological NFL state manifests unique electronic polarization characteristics akin to those in the non-interacting SSH model. Through exact diagonalization of the interacting SSH-HK Hamiltonian, we explore the spectral functions and density of states, revealing significant departures from traditional quasiparticle behavior in various particle number sectors. Our findings extend the understanding of topological non-Fermi liquids and their potential implications for high-temperature superconductivity and other correlated electron systems, highlighting the intricate interplay between topology and strong electron correlations.
\end{abstract}

\maketitle
\section{Introduction}

One of the primary challenges in condensed matter physics is elucidating the electric and magnetic properties of strongly correlated systems, where intricate interactions among numerous electrons complicate the determination of their ground state properties. Fermi liquid theory, introduced by Landau in 1963 \cite{landau1957theory}, serves as a foundational framework for understanding these systems, modeling interacting fermions as non-interacting quasiparticles. This theory begins with a non-interacting Fermi gas, progressively incorporating interactions, leading to an adiabatic transformation of the ground state into that of the interacting system.

Fermi liquid theory has proven remarkably successful in explaining the behavior of a wide variety of metals and electron systems at low temperatures. However, its limitations become apparent in strongly correlated systems, where interactions are so intense that the simple picture of non-interacting quasiparticles breaks down. Examples include high-temperature superconductors\cite{castellani1996non}, heavy fermion systems\cite{yu2021theoretical}, and certain magnetic materials\cite{maple1994non}. In these cases, phenomena such as non-Fermi liquid (NFL) behavior, emergent excitations, and complex ordering patterns arise, challenging our understanding. NFLs are characterized by excitations that are no longer long-lived quasiparticles, deviating from the expected behavior seen in non-interacting systems.

Recent studies have identified a novel class of NFLs with topological characteristics. For instance, in skyrmion crystals like MnSi, the interplay between conduction electrons and the collective excitations of the skyrmion lattice gives rise to a NFL phase, marked by anomalous temperature-dependent resistivity scaling as $\sim T^{\frac{2}{3}}$ \cite{ritz2013formation}. Additionally, topological NFLs have been observed in materials with strong spin-orbit coupling, such as pyrochlore iridates and the inverted band gap semiconductor ${\rm HgTe}$ \cite{moon2013non}. In these systems, Coulomb interactions facilitate the emergence of a topological NFL phase, which remains stable under the preservation of time-reversal and/or cubic symmetries.

In this study, we explore the emergence of a topological NFL phase within the Su–Schrieffer–Heeger (SSH) chain model subjected to long-range interactions. The SSH chain is recognized as a one-dimensional topological insulator and plays a pivotal role in elucidating the enhanced electrical conductivity observed in doped polyacetylene polymers. Our comprehensive analysis of the ground state phase diagram reveals that long-range interactions induce the manifestation of topological NFL characteristics over a substantial region of the phase diagram.

Recent investigations have increasingly scrutinized both topological and nontopological properties of SSH models, encompassing both $n$-dimensional structures \cite{liu2023analytic and one-dimensional chains}, particularly in the context of long-range hopping terms \cite{ahmadi2020topological, li2014topological} and spin-orbit couplings \cite{ vayrynen2011chiral, yan2014topological, bahari2016zeeman, yao2017theory}, as well as onsite Hubbard interactions \cite{le2020topological}, extended attractive and repulsive Hubbard interactions \cite{xing2023attractive, feng2022phase, cai2022robustness}. In the present work, we investigate the phase diagram of the SSH chain under the influence of the Hatsugai-Kohmoto (HK) interaction, characterized by its long-range behavior in real space and momentum-space localization. The HK interaction serves as the momentum-space analog of the Hubbard interaction \cite{hatsugai1992exactly}; unlike the onsite Hubbard interaction, which is localized in real space, the HK interaction's momentum-space localization facilitates the exact solvability of the HK model across all dimensions \cite{hatsugai1992exactly, zhao2024topic}.

Among the exactly solvable many-body systems exhibiting a non-Fermi liquid phase, the Sachdev-Ye-Kitaev model stands out due to its highly entangled ground state and non-trivial infrared dynamics, characterized by the absence of quasiparticles and properties fundamentally distinct from those of conventional Fermi liquids \cite{rosenhaus2019introduction, maldacena2016remarks}. The HK model has been utilized to explore the physical properties of various systems, including the coupling between localized moments and correlated electrons in heavy fermion cuprates such as $\mathrm{Nd_{x-1}Ce_xCuO_2}$ \cite{setty2021dilute}. It also addresses the NFL metal phase in cuprate superconductors \cite{phillips2020exact} and the two-stage superconductivity observed at lower temperatures \cite{li2022two}. Notably, the Fermi arcs in the HK model's ground state and excitation spectrum draw parallels to the phenomenology of high-temperature cuprate superconductors \cite{yang2021exactly}.

Recent studies have examined superconducting properties in a doped Mott insulator using the HK model, revealing significant deviations from standard BCS theory due to Mott physics. These findings include a first-order superconducting transition, a tricritical point, enhanced condensation energy, non-universal heat capacity jumps, and the suppression of the Hebel-Slichter peak in NMR, which may have implications for cuprates \cite{zhao2022thermodynamics}. Furthermore, it has been demonstrated that HK interactions can transform a topologically trivial superconductor into a nontrivial one, even in the absence of spin-orbit coupling \cite{zhu2021topological}.

To study quantum states in $f$-electron compounds like $\rm {SmB_6}$ and $\rm {YbB}_{12}$, researchers propose a modified periodic Anderson model that incorporates HK interactions \cite{zhong2022solvable}, recognizing that while the single band Hubbard model is useful for high-$T_c$ cuprates \cite{scalapino2012common, anderson1987resonating}, it lacks exact solutions in three-dimensional systems and necessitates approximate methods for comprehensive analysis.

In this paper, we analyze the SSH chain under the influence of the HK interactions (SSH-HK model) across various particle number sectors. We demonstrate that the ground state of the SSH-HK model exhibits NFL phases, distinguished by many-body Zak phases of $2\pi$ and $0$ for topological and trivial NFL phases, respectively. We show that the many-body Zak phase in the topological NFL state directly measures electronic polarization in the SSH-HK model, mirroring behavior in the non-interacting SSH model. Additionally, we explore spectral functions and density of states, revealing distinct quasiparticle behavior in different particle number sectors for NFL phases.

Our paper is organized as follows. In Sec. \ref{sec: SSHHK}, we introduce the SSH chain with long-range interactions, exactly diagonalize the interacting SSH-HK model, and present its eigenenergies. We also examine the behavior of the ground state particle densities. In Sec. \ref{sec: NFL}, we demonstrate that the ground state of the SSH-HK model possesses a non-Fermi liquid phase by performing the Luttinger integral. In Sec. \ref{sec: Spectral-DOS}, we analyze the spectral function of the system in different particle number sectors and investigate the behavior of the density of states. In Sec. \ref{sec: TNFL}, we compute the many-body Zak phase and examine the polarization of the system in the topological NFL phase. In Sec. \ref{sec: Chemical}, we investigate the impacts of chemical potential, showing that its presence leads to the emergence of a region in the phase diagram characterized by particle densities below 1. Finally, in Sec. \ref{sec: Conclude}, we summarize our results.

\section{SSH chain with long-range interactions}\label{sec: SSHHK}

\begin{figure}[!h]
	\centering
	\includegraphics[scale=0.35]{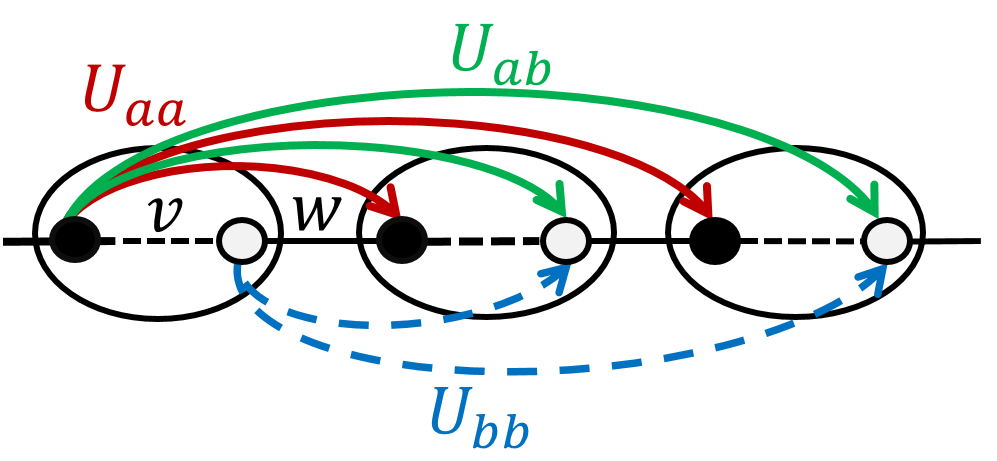}
	\caption{The schematic of a 1D SSH-HK model. The black solid (dashed) links represent the inter-(intra-)cell hopping energies $v (w)$. The black and white dots indicate the sublattices A and B within each unit cell. The green, red, and blue arrows illustrate the HK long-range interactions between two fermions located at A and B sites, respectively.} 
	\label{Fig:SchematicChain}
\end{figure}

The SSH chain is a one-dimensional (1D) topological insulator described by the following Hamiltonian \cite{su1979solitons}:
\begin{equation}\label{s1.1}
	H_{\rm SSH}= \sum_{j, \sigma} (v a_{j\sigma}^\dagger b_{j\sigma} +w a_{j+1\sigma}^\dagger b_{j\sigma}) + H.c.,
\end{equation}
where 
$a_{j\sigma}^\dagger$ 
$(a_{j\sigma})$
and
$b_{j\sigma}^\dagger$ 
$(b_{j \sigma})$ operators
create (annihilate) fermions $a$ and $b$ with spin $\sigma$ respectively at the A and B sites of $j$th unit cell.
Here, $v$ and $w$ represent the hopping energies within and between cells, respectively. Figure \ref{Fig:SchematicChain} illustrates the schematic representation of the SSH chain. This one-dimensional SSH chain is categorized within the BDI symmetry class of symmetry-protected topological states \cite{altland1997nonstandard}. Its ground state phase diagram reveals two distinct gapped phases: when $v<w$, the system is in a topologically non-trivial phase with a quantized Zak phase $Z=\pi$. At $v=w$, the energy gap closes, signaling a topological phase transition. For $v >w$, the energy gap reopens, yet a band inversion occurs, resulting in a transition to a topologically trivial phase with $Z=0$. 

Organic charge transfer salts exhibit fascinating electron-phonon interactions characterized by a coupling mechanism distinct from the conventional bond-length-dependent hopping framework. This alternative coupling arises from local electronic energy shifts induced by on-site displacements, as elucidated by the Holstein model \cite{holstein1959studies}. Such interactions can precipitate Peierls instability, resulting in a modulation of charge density and an accompanying energy gap. The dynamics of these systems can be effectively encapsulated by the Aubry-André Hamiltonian \cite{aubry1980analyticity}:
\begin{eqnarray}
\nonumber H_{\rm AA}&=&\sum_{i,\sigma} [t(a_{i \sigma}^\dagger b_{i \sigma} + a_{i+1 \sigma}^\dagger b_{i \sigma} + H.c.)\\
&&+ \Delta(a_{i \sigma}^\dagger a_{i \sigma} - b_{i \sigma}^\dagger b_{i \sigma})].
\label{HAA}
\end{eqnarray}
In this Hamiltonian \(H_{\rm AA}\), both parity $P$ and charge conjugation $C$ symmetries are violated, although the model remains $CP$ invariant. The regime where $|\Delta|<2|t|$ corresponds to a topological ground state, whereas $|\Delta|>2|t|$ yields a topologically trivial ground state. Combining the two equations Eq. \eqref{s1.1} and Eq. \eqref{HAA} results in the Rice-Mele Hamiltonian \cite{rice1982elementary}, which can also be reformulated into a generalized SSH model.
 
In the presence of HK long-range interactions, the SSH Hamiltonian is modified to include additional interaction terms. The modified Hamiltonian can be expressed as:
\begin{equation}
\label{Eq:Hamilton-SSHHK} 
	H = H_{\rm SSH} + H_{\rm HK},
\end{equation}
where
\begin{eqnarray}
	\nonumber H_{\rm HK} &=& \frac{U_{aa}}{N} \sum_{j_{1,\dots,4}\in A} \delta_{j_1 +j_3, j_2 + j_4} a_{j_1\uparrow}^\dagger a_{j_2\uparrow} a_{j_3\downarrow}^\dagger a_{j_4\downarrow}  \\
 \nonumber &+& \frac{U_{bb}}{N} \sum_{j_{1,\dots,4} \in B} \delta_{j_1 +j_3, j_2 + j_4} b_{j_1\uparrow}^\dagger b_{j_2\uparrow} b_{j_3\downarrow}^\dagger b_{j_4\downarrow}\\
\nonumber & +& \frac{U_{ab}}{N} \sum_{j_{1,2 \in A},~ j_{3,4 \in B}} \delta_{j_1 +j_3, j_2 + j_4} a_{j_1\uparrow}^\dagger a_{j_2\uparrow} b_{j_3\downarrow}^\dagger b_{j_4\downarrow} .
\end{eqnarray}
Here, $N$ is the number of unit cell.
The Kronecker delta function $\delta_{j_1 + j_3, j_2 + j_4}$ in the summations implies that the sums are taken over sites satisfying the condition $x_{j_1} + x_{j_3} = x_{j_2} + x_{j_4}$. This condition signifies that the center of mass of two particles is conserved during the scattering process. In the first (second) sum, the center of mass of two $a$ ($b$) particles is conserved, while in the third sum, this condition ensures the conservation of the center of mass of two $a$ and $b$ particles. The terms $U_{aa}$, $U_{bb}$, and $U_{ab}$ represent the interaction strengths between two $a$ particles, two $b$ particles, and one $a$ and one $b$ particle, respectively.
To determine the energy spectrum of the interacting system described above, we perform a Fourier transform on the Hamiltonian in Eq. \eqref{Eq:Hamilton-SSHHK} as follows:
\begin{equation}
\label{Eq:Hamilton-Kspace}
H =\sum_{k}H_k,
\end{equation}
where $H_k=H_{\rm SSH}(k) +H_{\rm HK}(k)$ with
\begin{align}
\nonumber	H_{\rm SSH}(k) &=\sum_{\sigma} \epsilon_{k} a_{k\sigma}^\dagger b_{k\sigma}+h.c.,  \\ 
\nonumber	H_{\rm HK}(k) &= U_{aa} a_{k\uparrow}^\dagger a_{k\uparrow} a_{k\downarrow}^\dagger a_{k\downarrow} + U_{bb} b_{k\uparrow}^\dagger b_{k\uparrow} b_{k\downarrow}^\dagger b_{k\downarrow} \\
\nonumber&+ U_{ab} a_{k\uparrow}^\dagger a_{k\uparrow} b_{k\downarrow}^\dagger b_{k\downarrow},
\end{align}
where $\epsilon_{k}=v+we^{ik}$. In the following, we consider a symmetric case where $U_{aa}=U_{bb}=U_{ab}=U$.
The long-range HK interaction is localized in momentum space, meaning that the interaction part of our Hamiltonian can be diagonalized through a Fourier transformation. To find the eigenenergies and eigenstates of the total Hamiltonian, we construct the matrix of the Bloch Hamiltonian $H_k$ in the basis:
\begin{align}\label{Eq:bases}
\ket{n_1, n_2, n_3, n_4}\equiv (a_{k\uparrow}^\dagger)^{n_1} (a_{k\downarrow}^\dagger)^{n_2} (b_{k\uparrow}^\dagger)^{n_3} (b_{k\downarrow}^\dagger)^{n_4} \ket{{\mathbf 0}},
\end{align}
where $n_i =0, 1$, and $\ket{{\mathbf 0}}=\ket{0,0,0,0}$ is the vacuum state, defined such that $a_{k\uparrow}a_{k\downarrow}b_{k\uparrow}b_{k\downarrow}\ket{{\mathbf 0}}=0$, indicating the absence of any fermions. The Bloch Hamiltonian conserves particle number, resulting in a block diagonal matrix in these bases. In the state $\ket{n_1, n_2, n_3, n_4}=\ket{n_1(k), n_2(k), n_3(k), n_4(k)}$, $n_{1(2)}$ and $n_{3(4)}$ represent the number of $a_{\uparrow (\downarrow)}$ and $b_{\uparrow (\downarrow)}$ fermions occupying the $k$-state, respectively. Each $k$ can be occupied by 0 to 4 particles in total. The 0-particle and 4-particle sectors each contain only one state, the 1-particle and 3-particle sectors each contain 4 states, and the 2-particle sector contains 6 states. In these bases, the Hamiltonian is a $16 \times 16$ block-diagonal matrix, consisting of 5 matrices: two $1 \times 1$ matrices, two $4 \times 4$ matrices, and one $6 \times 6$ matrix. This can be written as $H_k = H^{(0)}_{1 \times 1}\oplus H^{(1)}_{4 \times 4} \oplus H^{(2)}_{6 \times 6} \oplus H^{(3)}_{4 \times 4} \oplus H^{(4)}_{1 \times 1}$, where the superscripts denote the particle-number sectors and the subscripts indicate the dimension of the block Hamiltonian in each particle-number sector.
\begin{figure}[h]
	\centering
	\includegraphics[scale=0.16]{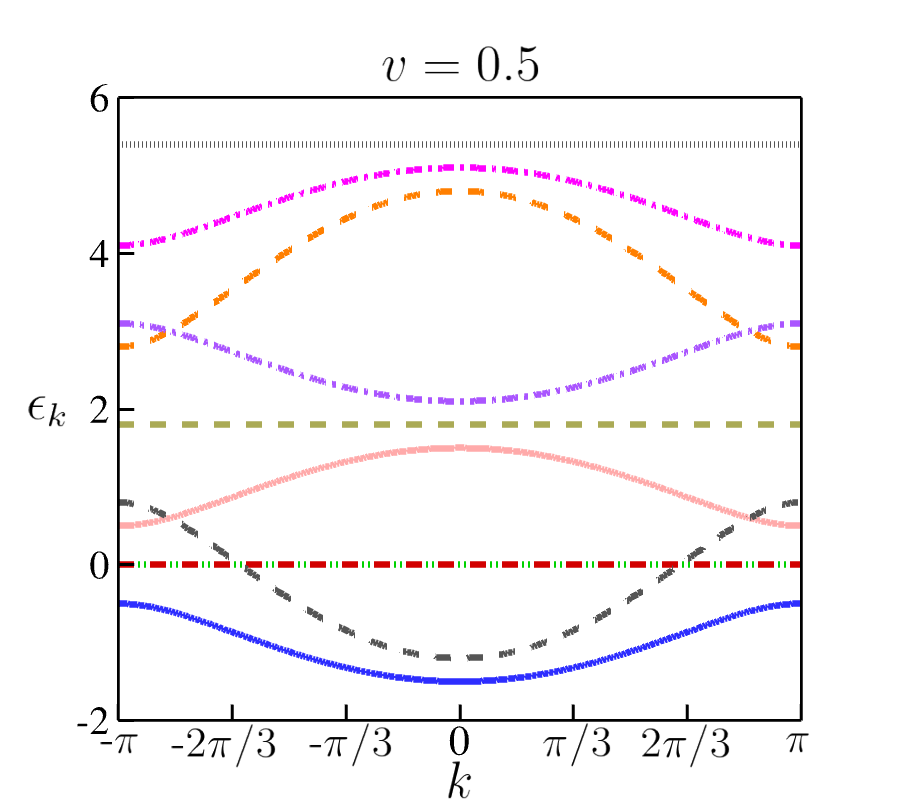}
	\includegraphics[scale=0.17]{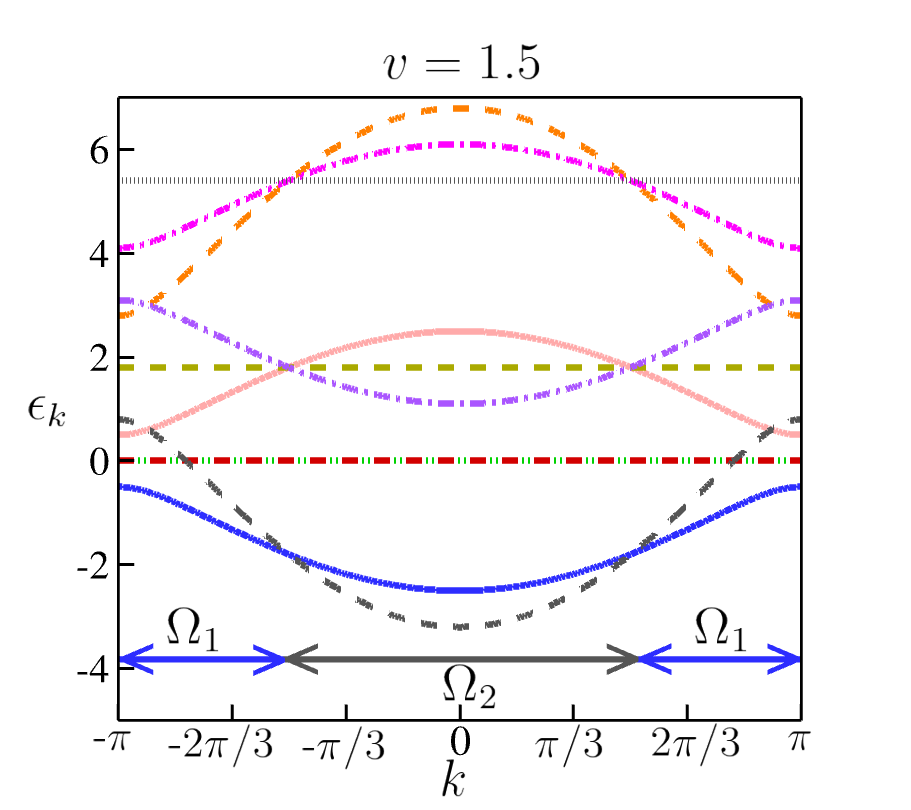}
	\includegraphics[scale=0.16]{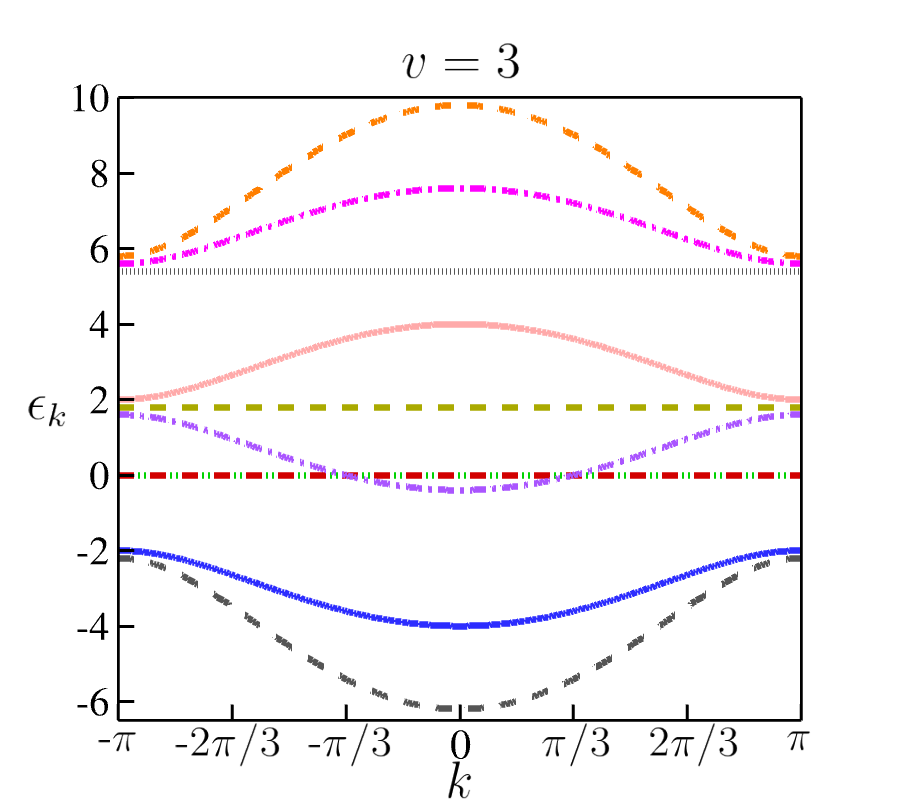}
	\caption{Eigenenergies of the SSH-HK Hamiltonian for $U=1.8$ with varying intracell hoppings. Top: $v=0.5$, middle: $v=1.5$, and bottom: $v=3$. In all plots the intercell hopping $w$ is set to 1. The dash-dot-dot line shows the 0-particle sector eigenenergy. Solid lines indicate the double degenerate eigenenergies of the 1-particle sector ($\pm|\epsilon_k|$). Dashed lines represent the 2-particle sector eigenenergies ($U$ and $U\pm2|\epsilon_k|$). Dash-dotted lines show the double degenerate eigenenergies of the 3-particle sector ($2U\pm|\epsilon_k|$). The dotted line represents the 4-particle sector eigenenergy $3U$. In the top and bottom cases, the ground state is confined to either the 1-particle or 2-particle sector. However, in the middle case, the ground state is composed of two distinct regions: in the vicinity of $k=0$ (designated as $\Omega_2$), it resides in the 2-particle sector, while in the surrounding area (denoted as $\Omega_1$), it is situated in the 1-particle sector. 
	}
\label{Fig:eigenenergy}
\end{figure}

The vanishing act of the Bloch Hamiltonain $H_k$ on the state $\ket{0000}$ in the 0-particle sector, and the states $\ket{1010}$ and $\ket{0101}$ in the 2-particle sector, indicates that these states do not interact with the other states. As a result, they do not contribute to the eigenstates of the Hamiltonian $H_k$, and consequently, obtaining the energy spectrum is effectively reduced to the diagonalization of three $4\times4$ matrices.
In the 1-particle sector, the Hamiltonian is given by:
\begin{align}\label{Eq:1-particle-H}
	H_k^{(1)}= \begin{pmatrix}
		0& \epsilon_{k} &0 & 0 \\ \epsilon_{k}^{\ast} & 0 &0 &0\\0& 0&0&\epsilon_{k}\\0 &0&\epsilon_{k}^\ast&0
	\end{pmatrix},
\end{align}
which has two double degenerate eigenenergies: 
\begin{align}\label{eigenh1}
	\epsilon_{k,\pm}^{(1)}=\pm|\epsilon_k|,
\end{align}
with the following eigenstates:
\begin{align}\label{Eq:1-particle}
	\ket{\Psi_{k}^{(1)}}^\pm_1 =\frac{1}{\sqrt{2}} \begin{pmatrix}
		0\\0\\ \pm e^{i\phi_k}\\ 1
	\end{pmatrix},
	\ket{\Psi_{k}^{(1)}}^\pm_2 =\frac{1}{\sqrt{2}}\begin{pmatrix}
		\pm e^{i\phi_k}\\ 1\\0\\0
	\end{pmatrix}.
\end{align}
Here, $e^{i\phi_k}=\frac{\epsilon_k}{|\epsilon_k|}$.
In the 2-particle sector, the Hilbert space dimension is six. However, as discussed, the eigenenergies can be obtained by diagonalizing a reduced 4$\times$4 Hamiltonian. The diagonalization of the following Hamiltonian
\begin{align}\label{Eq:2-particle-H}
	H_k^{(2)}= \begin{pmatrix}
		U & \epsilon_{k}^\ast &-\epsilon_{k}^\ast&0 & 0&0 \\ \epsilon_{k}&U&0&\epsilon_{k}^\ast &0&0\\-\epsilon_{k}&0&U&-\epsilon_{k}^\ast&0&0 \\
		0&\epsilon_{k}&-\epsilon_{k}&U&0&0\\
		0&0&0&0&0&0\\
		0&0&0&0&0&0
	\end{pmatrix},
\end{align} 
gives a double degenerate flat-band eigenstates ($\ket{\Psi_k^{(2)}}_{1,2}$) with eigenenergy $\epsilon_k^{(2)}=0$, and two degenerate states with eigenenergies $\epsilon_{k}^{(2)}=U$ are:
\begin{align}\label{Eq:2-particle-1}
\ket{\Psi_k^{(2)}}_3 =\frac{1}{\sqrt{2}}\begin{pmatrix}
		-e^{-2i\phi_k}\\ 0\\0\\1\\0\\0
	\end{pmatrix},
	\ket{\Psi_k^{(2)}}_4 =\frac{1}{\sqrt{2}}\begin{pmatrix}
		0\\1\\ 1\\0\\0\\0
	\end{pmatrix},
\end{align}
and the eigenstates with dispersive energies:
\begin{align}\label{Eq:eigenval-p2-5-6}
\epsilon_{k,\pm}^{(2)}= U \pm2|\epsilon_k|.
\end{align} 
Corresponding eigenstates reads as:
\begin{align}\label{Eq:2-particle-2}
	\ket{\Psi_k^{(2)}}^{-}_5 =\frac{1}{2}\begin{pmatrix}
		e^{-2i\phi_k}\\ e^{-i\phi_k}\\-e^{-i\phi_k}\\1\\0\\0
	\end{pmatrix},
	\ket{\Psi_k^{(2)}}^{+}_6 =\frac{1}{2}\begin{pmatrix}
		e^{-2i\phi_k}\\ -e^{-i\phi_k}\\ e^{-i\phi_k}\\1\\0\\0
	\end{pmatrix}.
\end{align}
The 3-particle sector corresponds to the following block-diagonal Hamiltonian
\begin{align}\label{Eq:3-particle-H}
	H_k^{(3)}= \begin{pmatrix}
		2U & -\epsilon_{k}^\ast &0 & 0 \\ -\epsilon_{k} &2U&0 &0\\0& 0&2U&-\epsilon_{k}^\ast\\0 &0&-\epsilon_{k}&2U
	\end{pmatrix}.
\end{align}  
This Hamiltonian has two double-degenerate eigenstates with the following eigenenergies 
\begin{align}\label{eigenh3}
	\epsilon_{k,\pm}^{(3)}=2U\pm|\epsilon_k|.
\end{align}
The eigenstates corresponding to the eigenenergies $\epsilon_{k,+}^{(3)}$ and $\epsilon_{k,-}^{(3)}$ are respectively $\ket{\Psi_k^{(3)}}^+_{1,2}$ and $\ket{\Psi_k^{(3)}}^-_{1,2}$, given by:
\begin{align}\label{Eq:3-particle}
	\ket{\Psi_k^{(3)}}^\pm_1 =\frac{1}{\sqrt{2}} \begin{pmatrix}
		0\\0\\ \pm e^{-i\phi_k}\\ 1
	\end{pmatrix},
	\ket{\Psi_k^{(3)}}^\pm_2 =\frac{1}{\sqrt{2}}\begin{pmatrix}
		\pm e^{-i\phi_k}\\ 1\\0\\0
	\end{pmatrix}.
\end{align}
Finally, the last eigenenergy corresponds to the 4-particle sector and is given by $\epsilon_k^{(4)}=3U$.

The previously discussed 16 states are the eigenstates of the Hamiltonian $H_k$. We will henceforth represent these eigenstates by $\ket{\Psi_k (j)}$, with $j=1$ indicating the ground state and $j$ ranging from 2 to 16 corresponding to the first through the 15th excited states, respectively.

In the 1-, 2-, and 3-particle sectors, the energy spectra, as plotted in Fig. \ref{Fig:eigenenergy}, satisfy $\epsilon^{(i)}_k=\epsilon^{(i)}_{-k}$ for all $k$-values. This property indicates that the time-reversal symmetry of the SSH chain is preserved in the presence of HK interactions. In the 1-particle sector, the energy spectrum is also symmetric with respect to the $\epsilon_k=0$ point, indicating 
the charge-conjugation symmetry. Furthermore, the eigenenrgies satisfy $\epsilon^{(i)}_k=-\epsilon^{(i)}_{-k}$, indicating the system's chiral symmetry. 

In the other sectors, the interactions lead to the breaking of these symmetries. However, the modified Hamiltonian
\begin{align}
\nonumber\tilde{H} =&\sum_{k} H_{\rm SSH}(k) + U_{aa} \sum_{k}(n_{k \uparrow}^a- 1/2)(n_{k \downarrow}^a -1/2) \\
\nonumber&+ U_{bb} \sum_{k} (n_{k \uparrow}^b-1/2)(n_{k \downarrow}^b -1/2)\\
\label{htilde}&+ U_{ab}\sum_{k} n_{k \uparrow}^a n_{k \downarrow}^b,
\end{align}
is invariant under charge-conjugation transformation $a_{j \sigma} \rightarrow a^\dagger_{j \sigma}$ and $ a^\dagger_{j \sigma} \rightarrow a_{j \sigma}$.
This Hamiltonian differs from the original one only by the chemical potential
$\mu^\prime=-U_{aa}/2=-U_{bb}/2$ and an additional constant.

\begin{figure}[!h]
	\centering
	\includegraphics[scale=0.18]{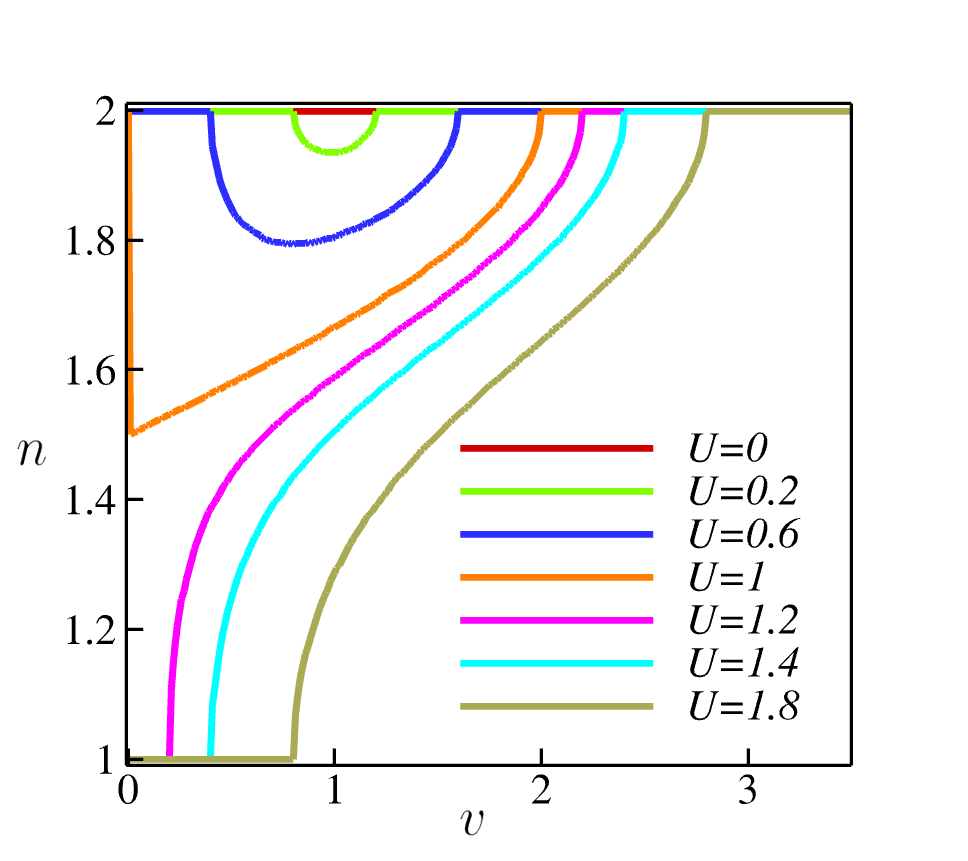}
	\includegraphics[scale=0.19]{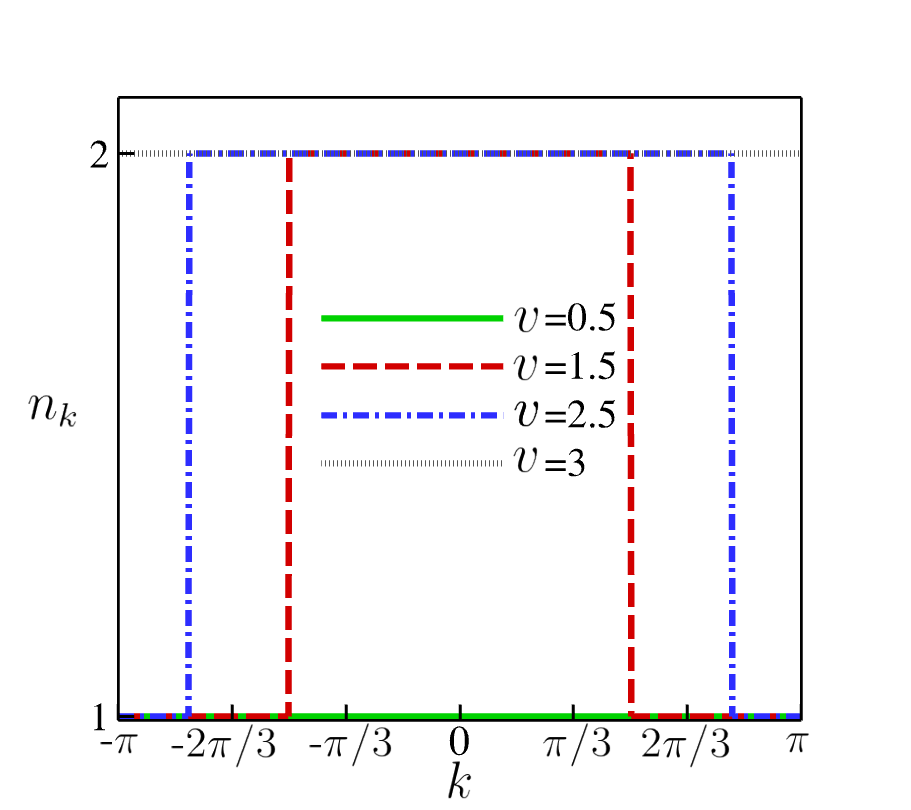}
	\caption{Top: Ground state particle density $n$ versus the intracell hopping amplitude ($v$) for various strengths of the interaction $U$. Here, the intercell hopping $w$ is set to $1$. Bottom: Plot of $n_k$ as a function of $k$ for different values of $v$ at $U=1.8$: For $v=1.5$ and $2.5$, a distinct surface delineates the regions of $n_k=1$ and $n_k=2$.} 
	\label{Fig:n-vs-v}
\end{figure}

The ground state of the Hamiltonian $H$ is a tensor product of the ground state of $H_k$. In some cases, the ground state of $H_k$ consists of a single band, as illustrated in Fig. \ref{Fig:eigenenergy} for $v=0.5$ and $v=3$. When the ground state consists of a single band, the system lacks both a Fermi surface, which distinguishes empty states from fully filled states, and a Fermi-like surface, which separates states with different fillings. In other cases, the ground state is composed of two or more bands that anticross at various $k$ points, as shown in the plot for $v=1.5$ in Fig. \ref{Fig:eigenenergy}. This results in a Fermi-like surface that separates single-occupied states from double-occupied states. To analyze the ground state of the system and its characteristics, we examine the behavior of the ground state particle density $n$, defined in terms of $a$ and $b$ particles distribution functions, as $n=n^a+n^b=\frac{1}{N}\sum_{k} (n_k^a+n_k^b)$. Here,
\begin{eqnarray}
&&\nonumber n^a_k= \frac{1}{N_d}\sum_{d}\sum_{\sigma} \bra{\Psi_{k}^d (1)}  a_{k\sigma}^{\dagger} a_{k\sigma} \ket{\Psi_{k}^d(1)},\\
&&n^b_k= \frac{1}{N_d} \sum_{d}\sum_{\sigma} \bra{\Psi_{k}^d(1)}  b_{k\sigma}^{\dagger}b_{k\sigma} \ket{\Psi_{k}^d(1)},
\label{s1.5}
\end{eqnarray}
with $\ket{\Psi_{k}^d(1)}$ representing the ground state of the Bloch Hamiltonian $H_k$. The sum $\frac{1}{N_d} \sum_{d}$ accounts for the degenerate ground states $\ket{\Psi_k^d(1)}$ with eigenenergy $E_k(1)$, and it is automatically excluded if the ground state is non-degenerate.
$N_d$ is number of degenerate states. In Fig. \ref{Fig:n-vs-v}-top, we have depicted the plots of $n$ as a function of $v$ for various strengths of $U$. Three distinct behaviors are highlighted:

\begin{enumerate}
\item{When the interaction $U$ is turned off, the particle density remains saturated at $n=2$ regardless of the value of $v$ (see the red line in Fig. \ref{Fig:n-vs-v}). In this scenario, the ground state of the system is a tensor product of the degenerate ground states of $H_k^{(2)}$, denoted by $\ket{\Psi_{k}^{(2)}}_{4}$. In these states, $ n^a = n^b=1$. Consequently, each $k$ state is, on average, fully occupied by $a_{\uparrow}$ and $a_{\downarrow}$, and $b_{\uparrow}$ and $b_{\downarrow}$ particles.}

\item{When the interaction $U$ is turned on, there is an interval around $v=1$ where the particle density drops below 2 (see the green and blue curves in Fig. \ref{Fig:n-vs-v}). Within this interval, near the $k=0$ point, the ground state of the $H_k$ is in the 2-particle sector, given by the states  $\ket{\Psi_{k}^{(2)}}_4$. However, away from this point, level crossings occur at two $k$ points symmetrically distributed around the $k=0$ point, and the ground state of the $H_k$ shifts to the 1-particle sector, represented by $\ket{\Psi_{k}^{(1)}}^-_{1,2}$. In the ground states corresponding to the 1-particle sector, $ n^a = n^b=\frac{1}{2}$, and each $k$ state is, on average, fully occupied by $a_{\uparrow}$ and $a_{\downarrow}$, and $b_{\uparrow}$ and $b_{\downarrow}$ particles.
However, in the ground state corresponding to the 2-particle sector, $n^a = n^b=1$. Consequently, each $k$ state is partially filled, leading to a ground state with a particle density of $n<2$.}

\item{As $U$ increases, the particle density exhibits different behaviors. When intracell hopping is small, the ground state particle density is $n=1$. This density increases with $v$ and eventually saturates at $n=2$ for $v > 1$. In the region with $n=1$, the ground state of the $H_k$ is in the 1-particle sector, given by the state $\ket{\Psi_{k}^{(1)}}^-_{1,2}$. In this state, 
$ n^a= n^b=\frac{1}{2}$, indicating that each $k$ state is, on average, half filled with $a_{\uparrow}$ and $a_{\downarrow}$, and $b_{\uparrow}$ and $b_{\downarrow}$ particles. The interaction $U$ prevents particles from occupying a single $k$ state, thus reducing the ground state particle density to $n=1$.
For small values of $U$, the ground state particle density is $n > 1$. However, when $U = 1$, a region with $n=1$ emerges and expands as the interaction strength increases.}
\end{enumerate}
In Fig. \ref{Fig:n-vs-v}-bottom, we present the plot of $n_k$ as a function of $k$ for various values of $v$ with $U=1.8$. The data reveals that the particle density $n_k=n_k^a+n_k^b$ exhibits a stepwise behavior, taking on values of either 1 or 2. However, when we consider the total sum over all $k$, a continuous transition from 1 to 2 is observed, as illustrated in Fig. \ref{Fig:n-vs-v}-top. Notably, in the presence of a chemical potential, $n_k$ remains stepwise but can also take on a value of zero, in addition to 1 and 2. This phenomenon results in Fermi-like surfaces that differentiate between the values of $n_k=0, 1$, and $2$, which we will explore further in section \ref{sec: Chemical}.

\section{Non-Fermi liquid phase}\label{sec: NFL}

In noninteracting free fermion systems, the concept of Fermi liquid theory is applicable, indicating the existence of a Fermi surface that delineates occupied from unoccupied quantum states. However, this theoretical framework becomes inadequate when interparticle interactions are considered. In 1960, Luttinger established that, with appropriate definitions and under specific conditions, a Fermi surface can be rigorously defined for interacting fermionic systems. The Luttinger integral provides a critical criterion, demonstrating that for interacting fermions, the Fermi volume, defined as the spatial volume enclosed by the Fermi surface, corresponds to the particle density as determined by the interacting Green's function. In such cases, the system retains the characteristics of a Fermi liquid. Conversely, a disparity between the Fermi volume and particle density identifies the system as a NFL. Luttinger's theory posits that when the particle density aligns with the Luttinger integral, the system is classified as a Fermi liquid. 

In our 1D system the Luttinger integral is defined as:
\begin{eqnarray}\label{Eq:Luttinger}
\nonumber&&I_a=\sum_{\sigma}\int \frac{dk}{2 \pi} \Theta(\Re G_\sigma^{aa} (k, \omega=0)),\\
&&I_b=\sum_{\sigma}\int \frac{dk}{2 \pi} \Theta(\Re G_\sigma^{bb} (k, \omega=0)),
\end{eqnarray}
where, $\Theta$ represents the heaviside step function, $\Re$ denotes the real part, and $G_\sigma^{aa}(k,\omega)$ and $G_\sigma^{bb}(k,\omega)$ are the ground state Green's functions of the particles $a_{\sigma}$ and $b_{\sigma}$. These functions are the Fourier transforms of $G^{aa}_\sigma(k, t) =-i\Theta (t)\langle{a_\sigma (k, t), a^\dagger_\sigma(k,0)}\rangle$,  $G^{bb}_\sigma(k, t) =-i\Theta (t)\langle{b_\sigma (k, t), b^\dagger_\sigma(k,0)}\rangle$. 
Using Lehmann representation (spectral representation) the Green's functions in Eq. \eqref{Eq:Luttinger} can be evaluated as\cite{coleman2015introduction}:
\begin{eqnarray}\label{Eq:Green-Lehman}
&&G_\sigma^{xy}(k,\omega)=\frac{1}{N_d} \sum_{d} \sum_{j=1}^{16}\\ 
\nonumber&&  [\frac{|\bra{\Psi_{k}^d(1)} x_{k \sigma} \ket{\Psi_{k}(j) }|^2 }{\hat{\omega}+ E_k(1) - E_k(j)} + \frac{|\bra{\Psi_{k}^d(1)}y_{k\sigma}^\dagger \ket{\Psi_{k}(j) }|^2 }{\hat{\omega}+ E_k(j) - E_k(1)}].
\end{eqnarray}
where, $x$ and $y$ indicates the $a$ or $b$ operators, and $\hat{\omega}=\omega+ i 0^+$. The state $\ket{\Psi_k(j)}$ with $j \neq 1$ denotes the $j$th excited state of the Bloch Hamiltonian $H_k$, with eigenenergy $E_k(j)$. 
The sum $\sum_{j=1}^{16}$ runs over all 16 eigenstates of the Hamiltonian $H_k$. 

\begin{figure}[!h]
	\centering
	\includegraphics[scale=0.175]{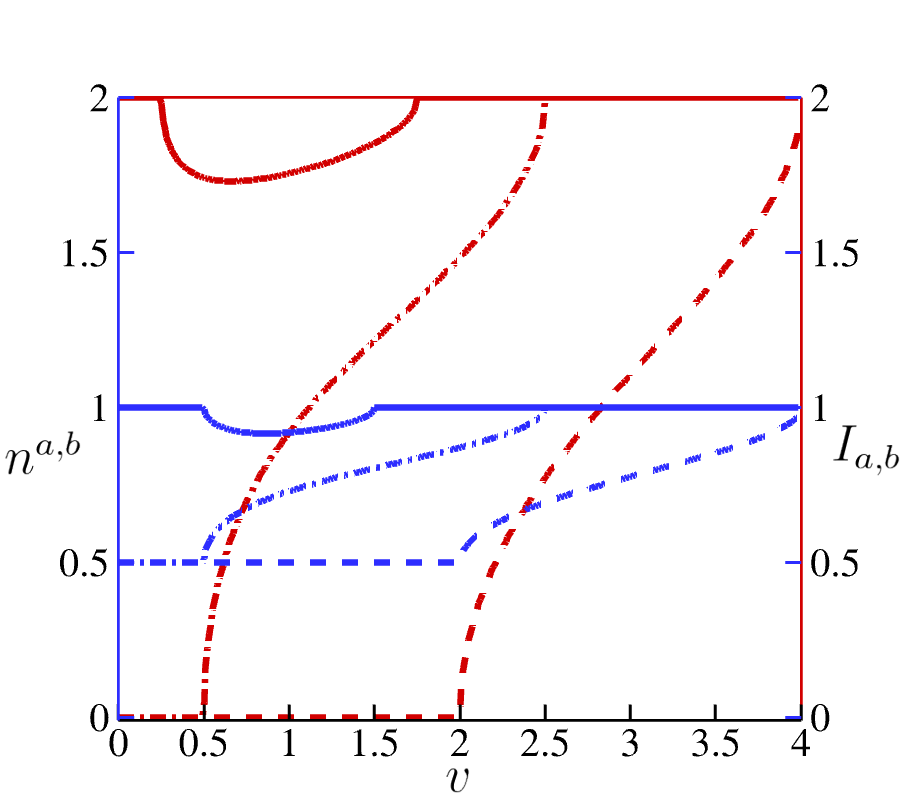}
	\caption{Ground state particle densities $n^a$ and $n^b$, along with the Luttinger integrals $I_a$ and $I_b$, as a function of the intracell hopping parameter $v$. The solid line represents $U = 0.5$, the dash-dot line corresponds to $U = 1.5$, and the dashed line indicates $U = 3$. Here, $n^a=n^b$, and $I_a=I_b$.
	}
	\label{Fig:LI-n}
\end{figure}

We have plotted in Fig. \ref{Fig:LI-n}, the Luttinger integrals, $I_a$ and $I_b$, along with the particle densities, $n^a$ and $n^b$, as a function of the intracell hopping parameter $v$, for various strengths of $U$. It is observed that for $U<1$, there is no value of $v$ where the particle densities match the Luttinger integrals, indicating that the entire phase diagram represents a NFL phase. On the other hand, when $U>1$, there exists at least one value of $v$ for each interaction strength at which the particle densities align with the Luttinger integrals, indicating that the ground state is characterized as a Fermi liquid. For instance, at $U=1.5$ (or 3), we find that for $v=0.7$ (or 2.3), the particle densities $n^{a,b}$ equal the Luttinger integrals $I_{a,b}$, confirming the Fermi liquid nature of the ground state.

\begin{figure}[!h]
	\centering
	\includegraphics[scale=0.19]{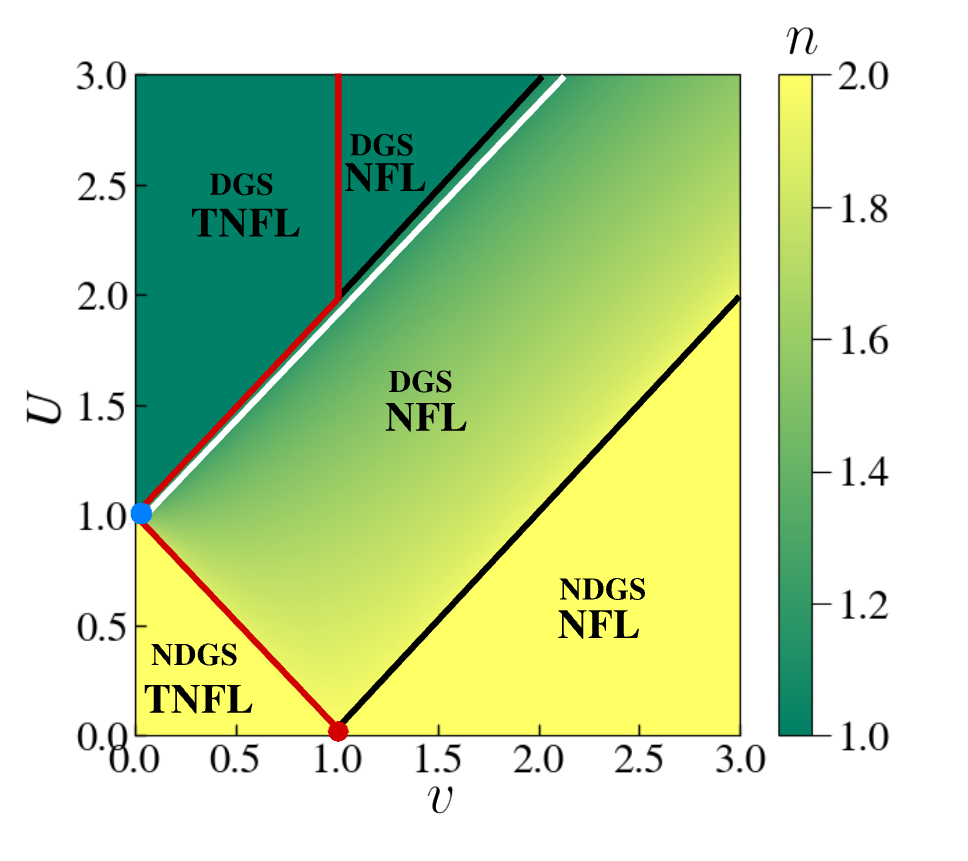}
	\caption{\small 
Ground state phase diagram of the SSH-HK model. The colors represent particle density. The ground state is a non-Fermi liquid in all regions except along the white line, where it is a Fermi liquid. Solid lines distinguish different non-Fermi liquid phases, with the red line separating the topological non-Fermi liquid (TNFL) phase from the trivial non-Fermi liquid (NFL) one. 'DGS' and 'NDGS' denote degenerate and nondegenerate ground states, respectively.
Along the black lines continuos phase transitions occur in the system. The blue dot at $U=1$ and $v=0$ marks a discontinuous phase transition between two topological NFL phases: the ground state is non-degenerate below $U<1$, and degenerate above $U>1$. The red dot represents the topological phase transition point of the SSH model ($U=0$), at half-filling ($n=2$). 
	} 
	\label{Fig:phase-diagram}
\end{figure}

In Fig. \ref{Fig:phase-diagram}, we have mapped out the $v-U$ ground state phase diagram of the SSH-HK model. The system is in the NFL phase throughout, except along the white line where the ground state is a Fermi liquid.

\section{Topological non-Fermi liquid}\label{sec: TNFL}

In the absence of interactions, the single particle SSH chain ($n=1$) exhibits a topological phase with a quantized Zak phase of $Z=\pi$, and a trivial phase with $Z=0$. When $v<1$ (with intercell hopping $w=1$), the SSH chain is gapped and is in the topological phase. At $v=1$ the energy gap closes, marking the topological-trivial phase transition. For $v>1$, the energy gap reopens with a band inversion, and the system transitions to the trivial phase. 

\subsection{many-body Zak phase}

Without interaction, the topological phase of the SSH chain is characterized by the Zak phase. When long-range HK interactions are present, the ground state is a NFL across most of the $v-U$ phase diagram, except along a line where it becomes a Fermi liquid (see the white line in the ground state phase diagram (Fig. \ref{Fig:phase-diagram})). To determine the topological properties of these NFL phases, we use the many-body Zak phase, defined as follows \cite{budich2013unraveling}:
\begin{align}\label{Eq:Zmb}
	Z_{mb}= i \int_{0}^{2 \pi} dk \bra{\Psi_{k}(1)} \partial_k \ket{\Psi_{k}(1)},
\end{align}
where $\ket{\Psi_k(1)}$ is the many-body ground state. This formula is used to examine the topological properties of nondegenerate ground states. In the phase diagram, the yellow regions indicate areas with a particle density of $n=2$, where the ground state is nondegenerate. 
In this region, the ground state is given by $\ket{\Psi_k^{(2)}}_5^-$ (see Eq.\eqref{Eq:2-particle-2}),
and the Zak phase simply reads:
\begin{eqnarray}
	\nonumber Z_{mb}&=& i \int_{0}^{2 \pi} dk [-i\partial_k\phi_k]\\
\label{Eq:Zmb-n2}&=&\oint d\phi_k= \Delta \phi_k=\begin{cases}
2 \pi &v < 1 \\ 0 & v > 1,
\end{cases}
\end{eqnarray}
where, $2\pi$ indicates a topological NFL phase, and $0$ indicates a topologically trivial NFL phase.

In the phase diagram, the green regions represent phases with particle density of $n=1$, where the ground state is doubly degenerate. To examine the topological properties of these phases, we use the following formula \cite{niu1985quantized}:
\begin{align}\label{Eq:deg-Zmb}
	Z_{dmb}=i \sum_d\int_{0}^{2 \pi} dk \bra{\Psi^d_{k}(1)} \partial_k \ket{\Psi^d_{k}(1)}.
\end{align}
Here, $\ket{\Psi^d_{k}}$ is the ground state, and $d$ denotes the ground state degeneracy. In this region, the degenerate ground states are given by $\ket{\Psi_k^{(1)}}^-_{1,2}$ (see Eq. \eqref{Eq:1-particle}), and the many-body Zak phase is:
\begin{eqnarray}
\nonumber Z_{dmb}&=& \frac i2 [\int_{0}^{2 \pi} dk (-i\partial_k\phi_k)+\int_{0}^{2 \pi} dk (-i\partial_k\phi_k)]\\
 \label{Eq:deg-Zmb-n1}&=&\Delta \phi_k=\begin{cases}
	2 \pi &v < 1 \\ 0& v > 1.
\end{cases}
\end{eqnarray}
This indicates that the NFL in the green region above the $v=1$ is trivial. For $v<1$, 
the many-body Zak phase is $2\pi$, indicating a topological NFL phase. 

In the region where $1<n<2$, we have a band-crossing between the degenerate and nondegenerate bands, and the ground state can be expressed as:
\begin{eqnarray}\label{GS_noninteger_n}
\ket{\Psi^d_{k}(1)}= \prod_{k \in \Omega_1} \ket{\Psi_k^{(1)}}_{1,2}^- \prod_{k \in \Omega_2} \ket{\Psi_k^{(2)}}_5^-,
\end{eqnarray}
where $\Omega_2$ designates a region surrounding $k=0$ in which the eigenstates $\ket{\Psi_k^{(2)}}_{5}^-$ represent the lowest energy state. Conversely, $\Omega_1$ comprises two regions adjacent to $\Omega_2$ where the states $\ket{\Psi_k^{(1)}}_{1,2}^-$ are degenerate eigenstates with the lowest energy. The regions $\Omega_1$ and $\Omega_2$ are depicted in Fig. \ref{Fig:eigenenergy}.
In this region, many-body Zak phase is not quantized, meaning that the NFL phase is topologically trivial.

\subsection{Polarization of the topological NFL}

The many-body Zak phase is a physically meaningful quantity. Similar to the Zak phase which directly provides the electronic polarization in crystalline solids\cite{resta1994macroscopic, king1993theory}, the many-body Zak phase also offers valuable information regardaing the electronic polarization of solids in the presence of electron-electron interactions. 
Following, we demonstrate that the many-body Zak phase derived above is directly proportional to the electronic polarization of the SSH-HK model.

In a noninteracting 1D system with periodic boundary condition the electronic contribution to the polarization of the system is\cite{king1993theory}:
\begin{align}
\label{eq: pol-el-NT1}P(\lambda) &=\frac{-e}{L} \sum_{nk}^{occ} \int_L r | \Psi_{nk} (r, \lambda)|^2 dr,
\end{align}
where $\Psi_{nk}(r,\lambda)$ is the wave function of electrons with wave number $k$ in the band $n$, $e$ is the absolute value of the electron charge, and $L$ is the system size. Here, the sum is over all the occupied bands, and the integral is evaluated over the system size. The parameter $\lambda$ is an external variable in the Hamiltonian, continuously varying between 0 and 1. For instance, in the 1D SSH model, the hopping parameters can be expressed in terms of an external variable as $v = (1 + \delta)t$ and $w = (1 - \delta)t$, with $\delta$ continuously varying between 0 and 1.
In the noninteracting case, the wavefunction is expressed in terms of Bloch wavefunction as $\Psi_{nk} (r, \lambda) = e^{ikr} u_{nk}(r, \lambda)$, and the polarization simplifies to
\begin{align}
\label{eq: pol-el-NT} P(\lambda)= \frac{-e}{L} \sum_{nk}^{occ}\bra{u_{nk}(\lambda)}\hat{r} \ket{u_{nk}(\lambda)}.
\end{align}
The quantity $P$ is ill defined because the diagonal matrix elements of the position operator are not well-defined for periodic wavefunctions. 
This issue is addressed by employing the relationship between the Zak phase and the derivative of $P$ with respect to $\lambda$, as follows\cite{king1993theory}: 
\begin{align}\label{eq:change-pol}
\frac{\partial P}{\partial \lambda}=\frac{-e}{L}\sum_{nk}^{occ} 2 \Im \left\langle {\partial_k u_{nk}(\lambda)}|\partial_\lambda u_{nk}(\lambda)\right\rangle
\end{align}
where $\Im$ denotes the imaginary part. Following detailed mathematical analysis, the polarization is determined as:
\begin{align}\label{pol-zak} 
	P(\lambda)= \frac{e}{2\pi} \sum_{nk}^{occ} \Im\int_{-\pi}^{\pi} dk \bra{ u_{nk}(\lambda)}  \partial_k \ket{u_{nk}(\lambda)}.
\end{align}

Therefore polarization of noninteracting systems is proportional to the non-interacting Zak phase as: $P= \frac{e Z}{2 \pi}$.

In interacting systems, the complexity increases markedly. Typically, the many-body Hamiltonian cannot be exactly diagonalized, requiring low-energy field theoretical approach \cite{di2024topological}, or numerical methods to diagonalize the Hamiltonian for a finite system size $L$. In this context, the system's eigenstate is represented in real space, and 
to examine topological propeties of the system, twisted boundary conditions $\Psi_\theta(r_j + L) = e^{i \theta} \Psi_\theta(r_j)$ are typically used. Consequently, the wave function is expressed as $\Psi_\theta = e^{i \theta \sum_{j=1}^{{\cal N}} r_j} \psi_\theta$, where ${\cal N}$ is the number of particles. Here,
 $\psi_\theta$ serves as the many-body counterpart to $u_k$ in the noninteracting case. The system is threaded with magnetic flux $\theta$, and the many-body Zak phase picked up by the many-body ground state when this magnetic flux is varied by one magnetic flux quantum is given by\cite{grusdt2013topological, resta1995many}:
\begin{align}\label{eq: many-body-zak-tb}
Z=i \int_{0}^{\frac{2\pi}{L} } d\theta \bra{\psi_\theta} \partial_\theta \ket{\psi_\theta}.
\end{align}
The many-body polarization is given in terms of the Zak phase as\cite{ortiz1994macroscopic}:
\begin{align}\label{eq: many-body-zak-tb}
P=\lim_{{\cal N},L\rightarrow \infty} \frac{e Z}{2\pi}.
\end{align}

In our exactly solvable interacting system, the interaction terms are localized in momentum space, making momentum a good quantum number. Despite the interactions, the system's eigenstates are determined exactly, as demonstrated, with the eigenfunctions expressed in terms of the Bloch Hamiltonian wave functions $H_k$ as $e^{ikr}\Psi_{k}(r)$. 
The procedure to obtain the polarization here is analogous to the noninteracting case, with the key difference being that in our interacting system, the ground state is not the occupied band but an eigenstate of the Bloch Hamiltonian $H_k$ with the minimum eigenenergy. Therefore, the polarization is given by:
\begin{align}
P_{mb}=\frac{-e}{L} \sum_k\bra{\Psi_{k}(1)}\hat{r} \ket{\Psi_{k}(1)},
\end{align}
and consequently
\begin{align}\label{eq:pol-many-body}
P_{mb}=\frac{i e}{2 \pi}\int_{0}^{2\pi} dk \bra{\Psi_{k}(1)} \partial_k \ket{\Psi_{k}(1)} = \frac{e Z_{mb}}{2 \pi}.
\end{align}

All physical quantities associated with polarization can be evaluated using many-body Zak phase. Experimentally, polarization changes in crystals can be induced by applying stress in piezoelectric materials, increasing temperature in pyroelectric materials, or reversing spontaneous polarization with electric fields in ferroelectric materials. 
In the topological NFL phase with $Z=2\pi$, charge polarization occurs with $P=e$, while trivial phases with a vanishing Zak phase exhibit no charge polarization. For degenerate ground state, a summation on degenerate states is required to obtain the correct Zak phase of the system. In this case the electronic polarization equals $\frac{e Z_{dmb}}{2 \pi}$.

\section{Spectral function and density of states (DOS)}\label{sec: Spectral-DOS}

The spectral function, defined in terms of the imaginary part of Green's function as $A(k,\omega)=-\frac{1}{\pi}\Im G(k, \omega)$, encodes information about the excitation spectrum of a system. For free non-interacting particles, the spectral function is a delta function at the pole of the Green's function, $A(k, \omega)=\delta(\omega - \xi_k)$, with $\xi_k$ representing the energy of the elementary excitation. In systems with interacting particles, the spectral function may deviate significantly from this simple form, but if it has a relatively narrow, well-defined peak, it can be interpreted as a quasiparticle with a finite lifetime, determined by the peak's width. 

The spectral function is a powerful tool for analyzing the electronic properties of interacting systems, particularly in strongly correlated regimes. In conventional Fermi liquids, quasiparticles are long-lived particels described by Landau's theory. Conversely, in NFLs, the quasiparticle paradigm often fails due to unconventional scattering processes and strong electronic correlations. This breakdown is typically manifested in broadening of spectral features, redistribution of spectral weight, or asymmetries in the spectral lineshape\cite{varma2002singular}.

Under specific conditions, NFLs can exhibit pronounced peaks in the spectral function, suggesting the emergence of coherent excitations that possess characteristics markedly different from those found in conventional Fermi liquids. These phenomena are often associated with interaction-driven mechanisms or the proximity to quantum critical points, where NFL behavior coexists with remnant quasiparticle-like attributes.

In the following, we investigate the behavior of the spectral function of the SSH-HK model. Notably, despite the system exhibiting a NFL phase, the spectral functions across different particle number sectors display Dirac delta form, indicating the presence of long-lived quasiparticles.
The positive-frequency part of $A(k,\omega)$ (for $\omega>0$) describes the process of adding an electron to the system, which is studied by inverse photoemission. For $\omega>0$, the spectral function $A(k,\omega)$ directly gives the intensity of the angle-resolved inverse photoemission (IPES) spectrum, indicating the probability of finding the system in a state with energy $\omega$ and momentum $k$ after adding a particle. Conversely, for $\omega<0$, $A(k,\omega)$ describes the probability of removing a particle, leaving the system in a state with energy $\omega$, which corresponds to angle-resolved photoemission (ARPES)\cite{zhang2023density, khomskii2010basic}.

Based on the Green's function in Eq. \eqref{Eq:Green-Lehman}, and the equation $A^{xy}(k, \omega)=\sum_{\sigma}A^{xy}_\sigma$, with $A^{xy}_\sigma=-\frac{1}{\pi}\Im G_\sigma^{xy}$, it is evident that all spectral functions: $A^{aa}(k,\omega)$, $A^{bb}(k,\omega)$, $A^{ab}(k,\omega)$, and $A^{ba}(k,\omega)$, are identical in our one-dimensional system. In the following, we remove the superscripts $a$ and $b$ and report the spectral function for each particle number sector separately.

In 1-particle sector with $n=1$, the ground state energy is $E_k(1)=|\epsilon_{k}|$, and the spectral function is expressed as follows: 
\begin{eqnarray}\label{eq: SF-of-excitations-n=1}
	A(k, \omega)&=&\frac{1}{2}\big[\delta(\omega-[U+|\epsilon_k|])+\delta(\omega-|\epsilon_k|)+\delta(\omega+|\epsilon_k|) \nonumber \\
	&&+\cos^2\phi_k \delta(\omega-[U-|\epsilon_k|]) \nonumber \\ 
	&&+\sin^2\phi_k \delta(\omega-[U+3 |\epsilon_k|])\big].
	\end{eqnarray}
This spectral function indicates that, within the NFL phase at the $n=1$ sector, there are five distinct excitations characterized by the dispersion energies $\omega_k^\pm=\pm|\epsilon_k|$, $\omega_k^\pm=U\pm|\epsilon_k|$, and $\omega_k=U+3|\epsilon_k|$. The normalization of this spectral function is given by $\frac{1}{2}\sum_\sigma\int d\omega A_\sigma(k, \omega)=1$.
\begin{figure}[h]
	\centering
		\includegraphics[scale=0.24]{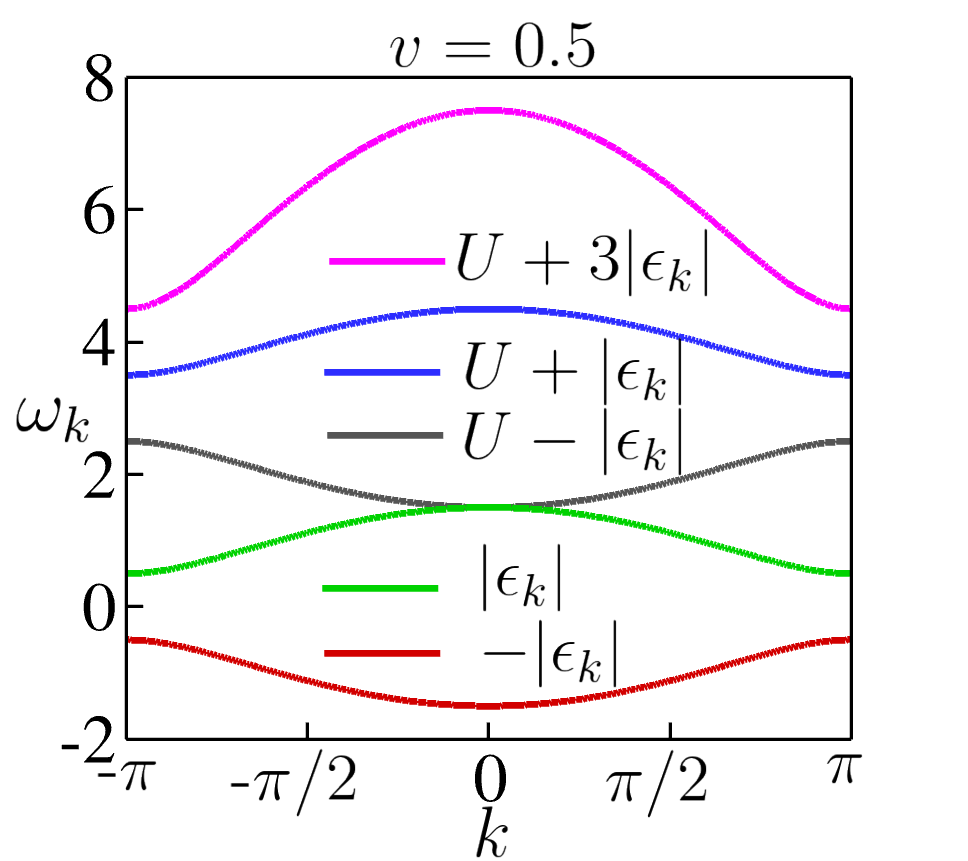}
		\includegraphics[scale=0.126]{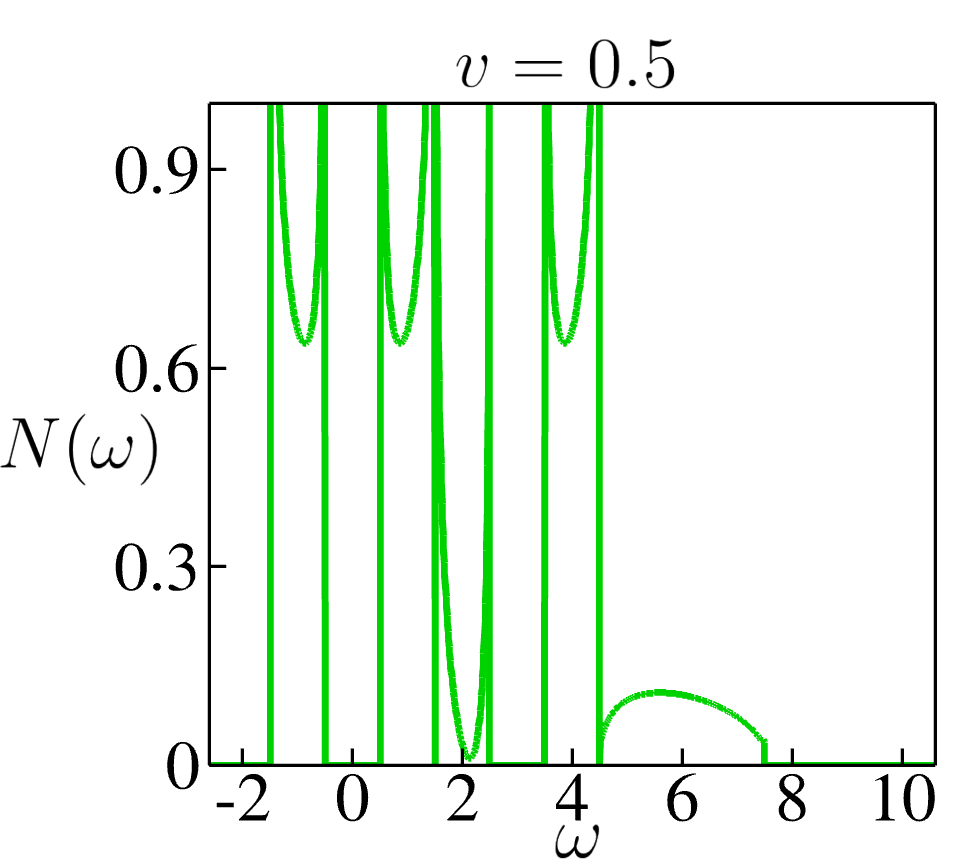}
		\includegraphics[scale=0.24]{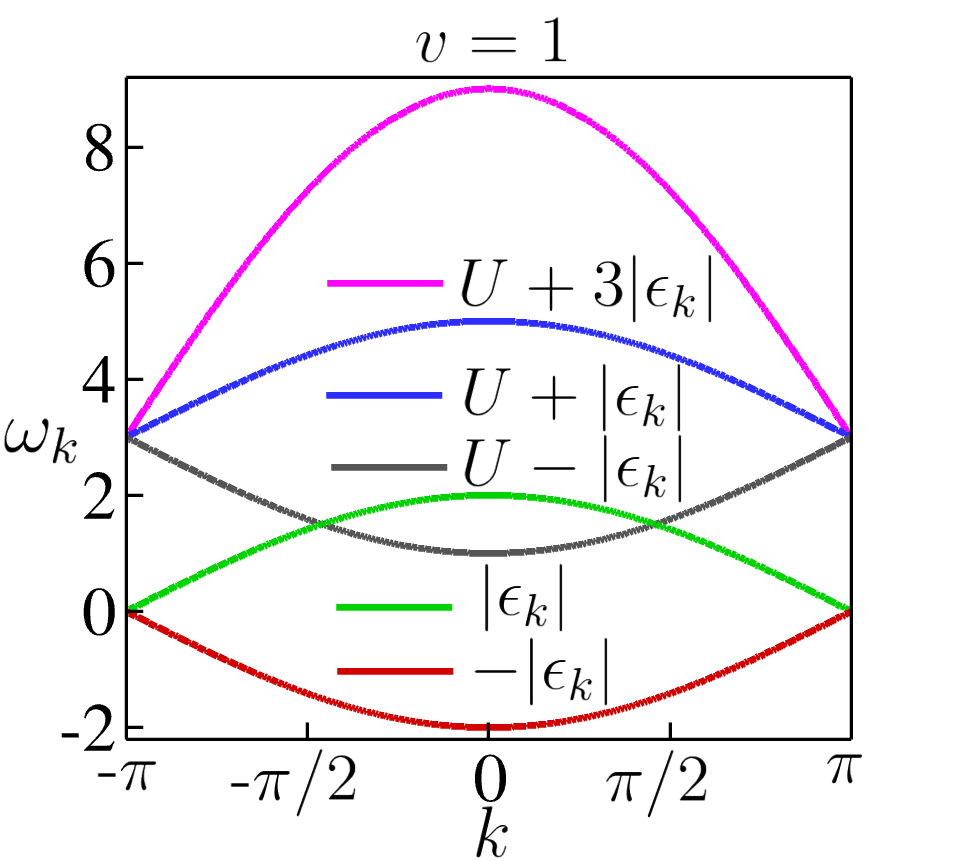}
		\includegraphics[scale=0.126]{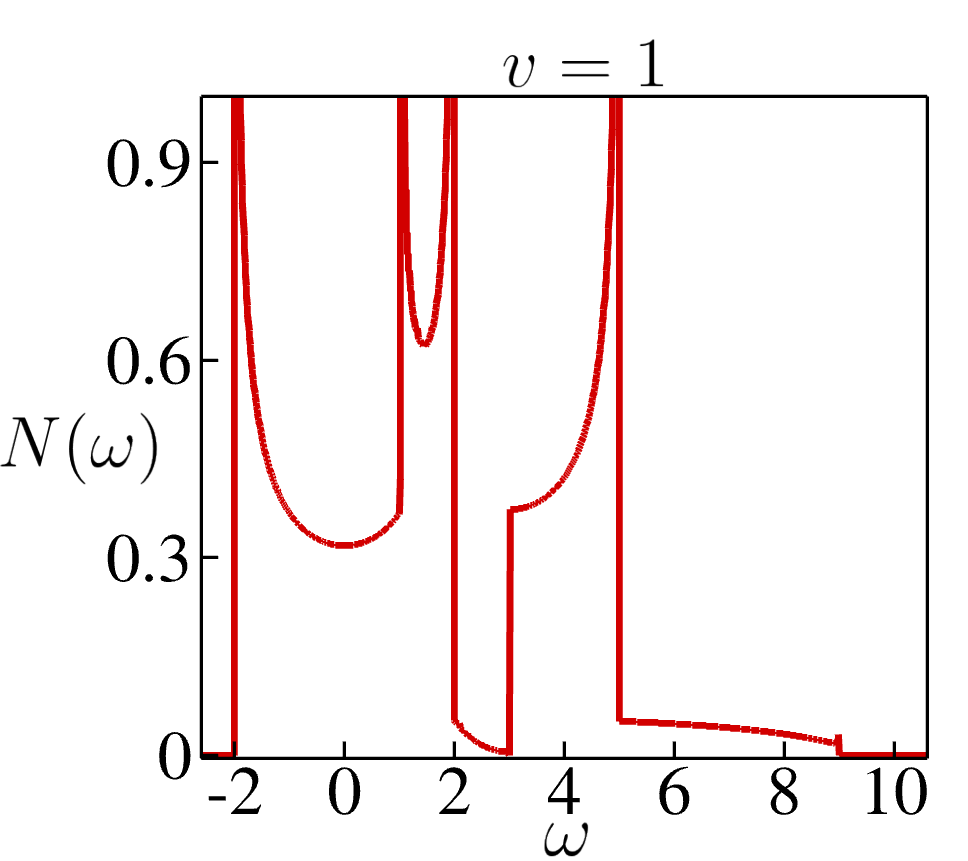}
		\includegraphics[scale=0.24]{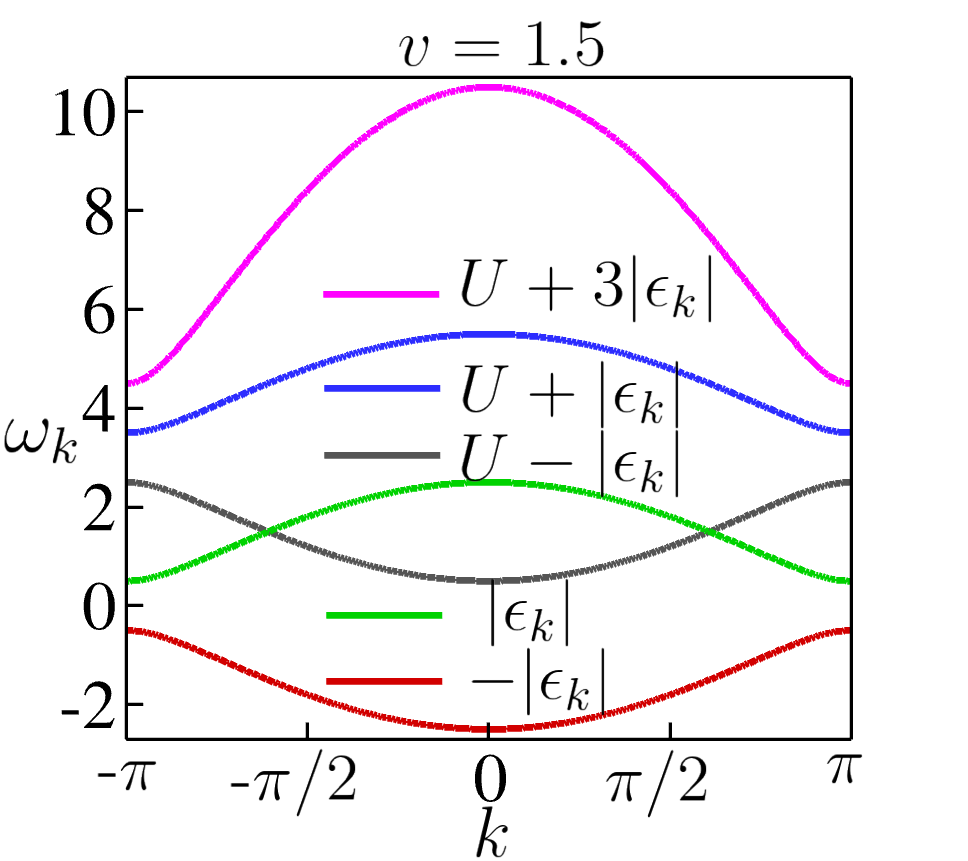}
		\includegraphics[scale=0.126]{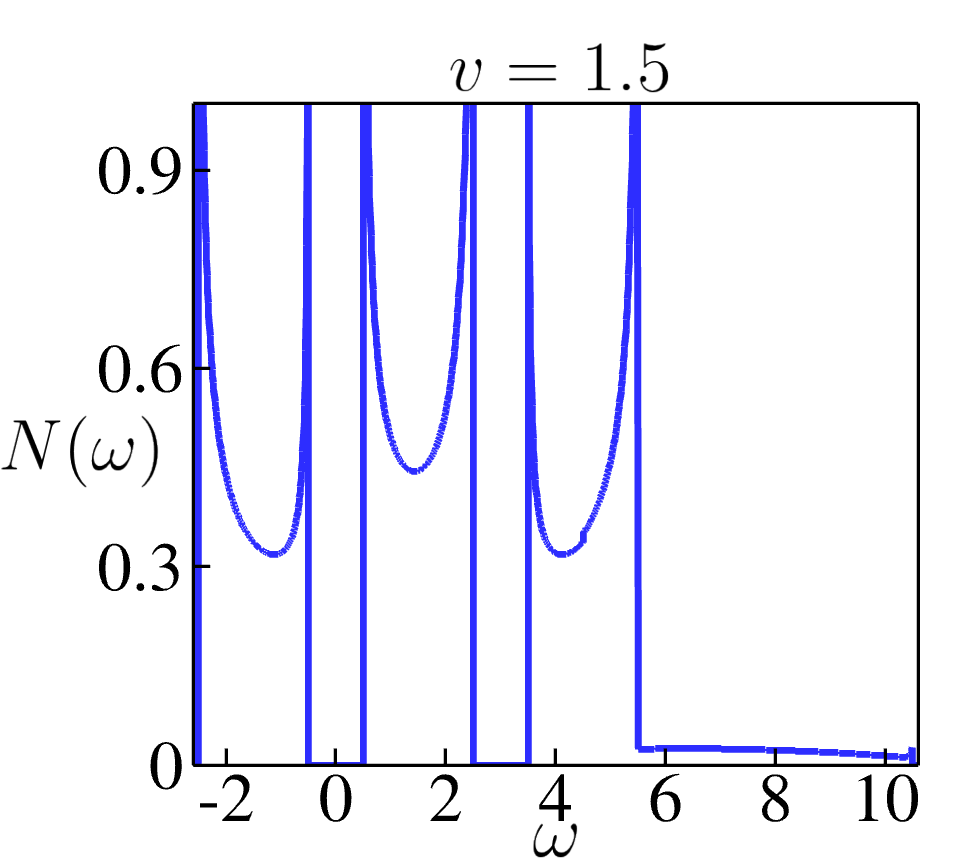}
		\caption{Energy of excitations (left column) and DOS (right column) in the 1-particle sector ($n=1$). All plots are generated with the HK interaction set to $U=3$, while the intra-cell hopping $v$ varies from $0.5$ to $1.5$. The excitation gap around $\omega_k=0$ closes at $v=1$, marking the transition of the system into a trivial phase. Here, the inter-cell hopping $w$ is set to 1.} 
\label{Fig: DOSn1}
\end{figure}

We present in Fig. \ref{Fig: DOSn1} the energy of excitations for varying strengths of the intracell hopping parameter $v$ in the 1-particle sector with $U=3$. 
For $v<1$, a gap is observed around $\omega_k=0$, which diminishes as $v$ increases and ultimately closes at $v=1$, marking a transition of the ground state into a trivial phase, and reopens for $v>1$. To investigate the impact of the interaction strength $U$, we analyzed the excitation energy bands at different values of $v$ (below, at, and above $v=1$). Our findings (not shown here) indicate that at any values of $v$ either below or above $1$, the excitation gap is present, with variations in $U$ affecting only the magnitude of the gap. Focusing on the transition point at $v=1$ and incrementing the strength of $U$ (along the red vertical line in Fig. \ref{Fig:phase-diagram}), we find that the system remains gapless until $U=U_c=4$. At this point, while the excitation gap around $\omega_k=0$ remains closed, a distinct energy gap emerges between two higher energy bands, which manifests in the optical properties of the system.

\begin{figure}[h]
	\centering
		\includegraphics[scale=0.24]{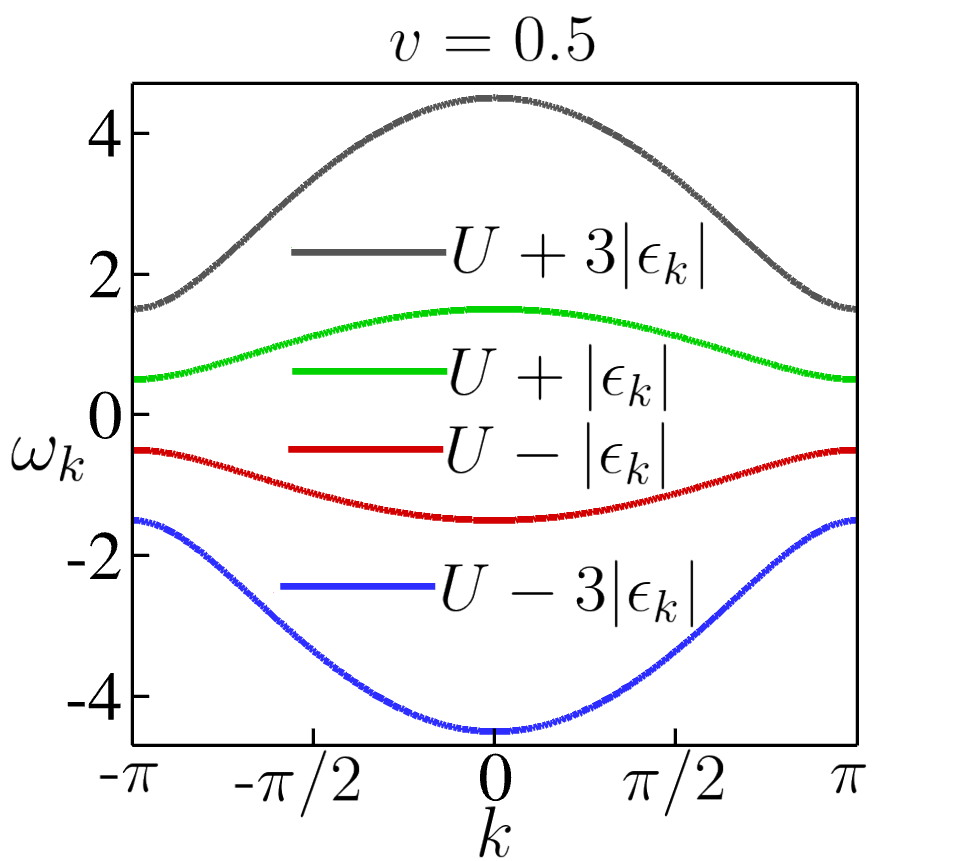}
		\includegraphics[scale=0.126]{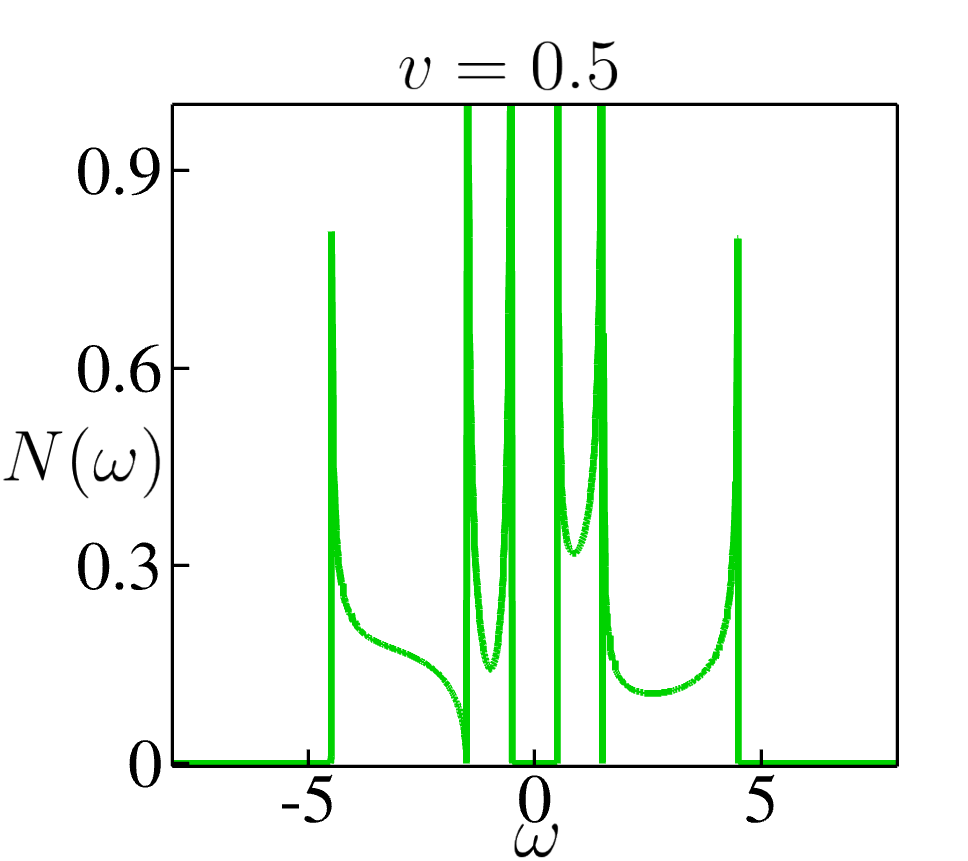}
		\includegraphics[scale=0.24]{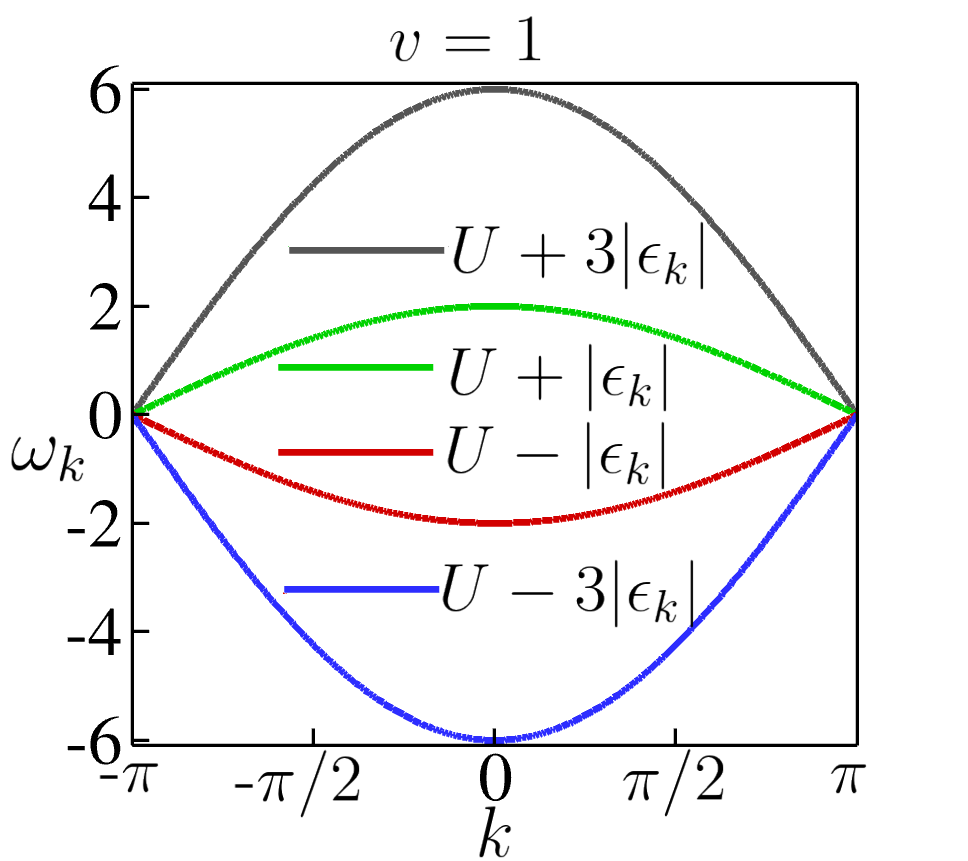}
		\includegraphics[scale=0.126]{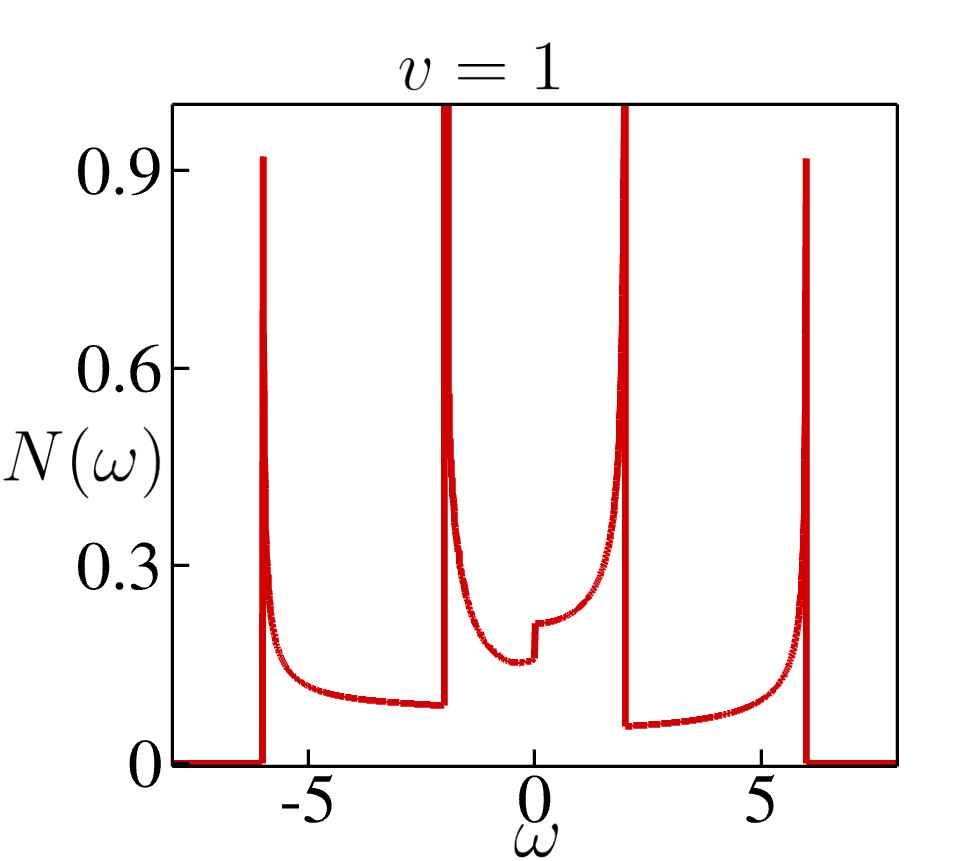}
		\includegraphics[scale=0.24]{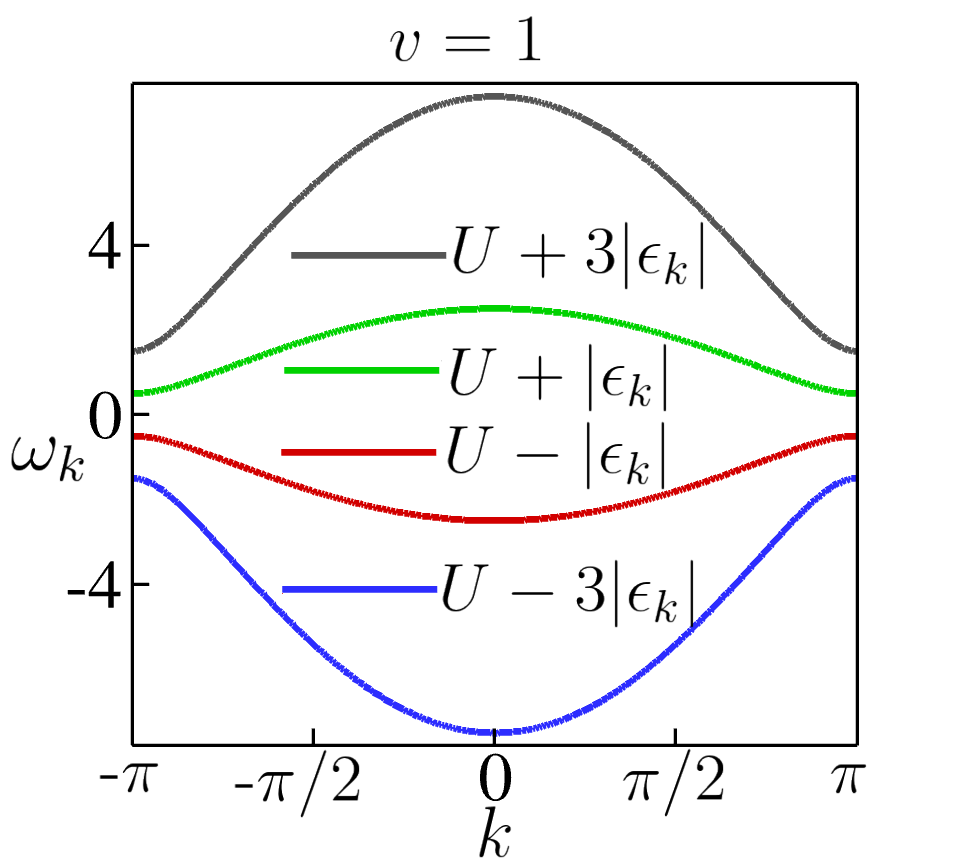}
		\includegraphics[scale=0.126]{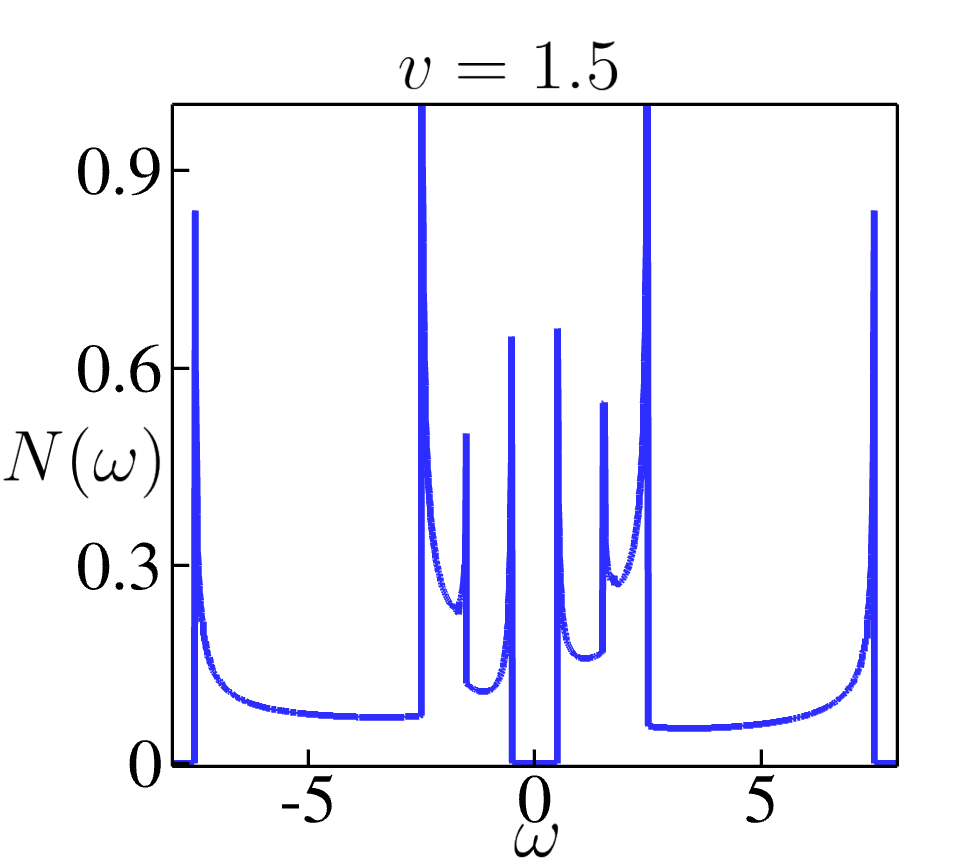}
		\caption{Energy of excitations (left column) and DOS (right column) in the 2-particle sector ($n=2$). All plots are generated with the HK interaction set to $U=0$, while the intra-cell hopping $v$ varies from $0.5$ to $1.5$. The excitation gap around $\omega_k=0$ closes at $v=1$, marking the transition of the system into a trivial phase. Here, the inter-cell hopping $w$ is set to 1.} 
\label{Fig: DOSn2}
\end{figure}
In 2-particle sector with $n=2$, the ground state energy is $E_k(1)=U-2|\epsilon_{k}|$, and the spectral function is expressed as follows: 
\begin{eqnarray}\label{eq: SF-of-excitations-n=2}
	&&A(k, \omega) =\frac{1}{2}\big[\delta(\omega-[U+|\epsilon_{k}|])+\delta(\omega-[U+3|\epsilon_{k}|]) \nonumber \\
	&&+(\cos^2\phi_k+\sin^2(\phi_k/2)) \delta(\omega-[U-|\epsilon_{k}|]) \nonumber \\ 
	&&+ (\sin^2\phi_k+\cos^2(\phi_k/2))\delta(\omega-[U-3|\epsilon_{k}|])\big].
\end{eqnarray}
This indicates that within the NFL phase in the 2-particle sector, there are four distinct excitations characterized by the dispersion energies $\omega_k^\pm=U\pm|\epsilon_k|$, and $\omega_k^\pm=U\pm3|\epsilon_k|$. These energies are illustrated in Fig. \ref{Fig: DOSn2}, for the specific case of $U=0$. The normalization of this spectral function is given by $\frac{1}{2}\sum_\sigma\int d\omega A_\sigma(k, \omega)=1$.

In the 2-particle sector, regardless of the values of $v$ (whether below or above 1), the excitation gap remains open as $U$ varies, resulting in the system acting as an insulator. Specifically, for $v<1$, it behaves as a topological insulator, while for $v>1$, it is classified as a trivial insulator. In this sector, at half-filling, the two energy bands below $\omega_k=0$ are fully occupied, whereas the two bands above this energy are unoccupied.

Within the 2-particle sector, at the critical point $v=1$ the excitation gap is zero, and increasing $U$ keeps the gap closed. This behavior differs from the standard HK model, which shows metallic properties for small $U$ relative to the bandwidth $W$, transitioning to a Mott insulating phase as $U$ approaches $W$, resulting in the closure of the excitation gap.

\begin{figure}[h]
\centering
		\includegraphics[scale=0.35]{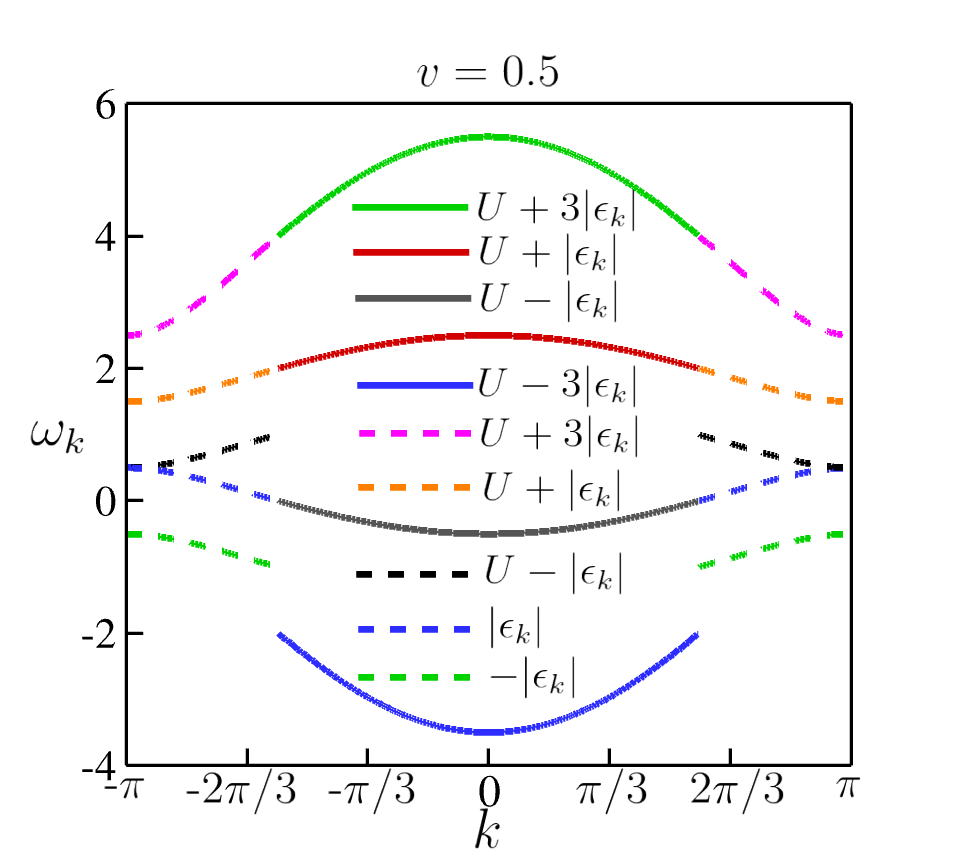}
		\includegraphics[scale=0.35]{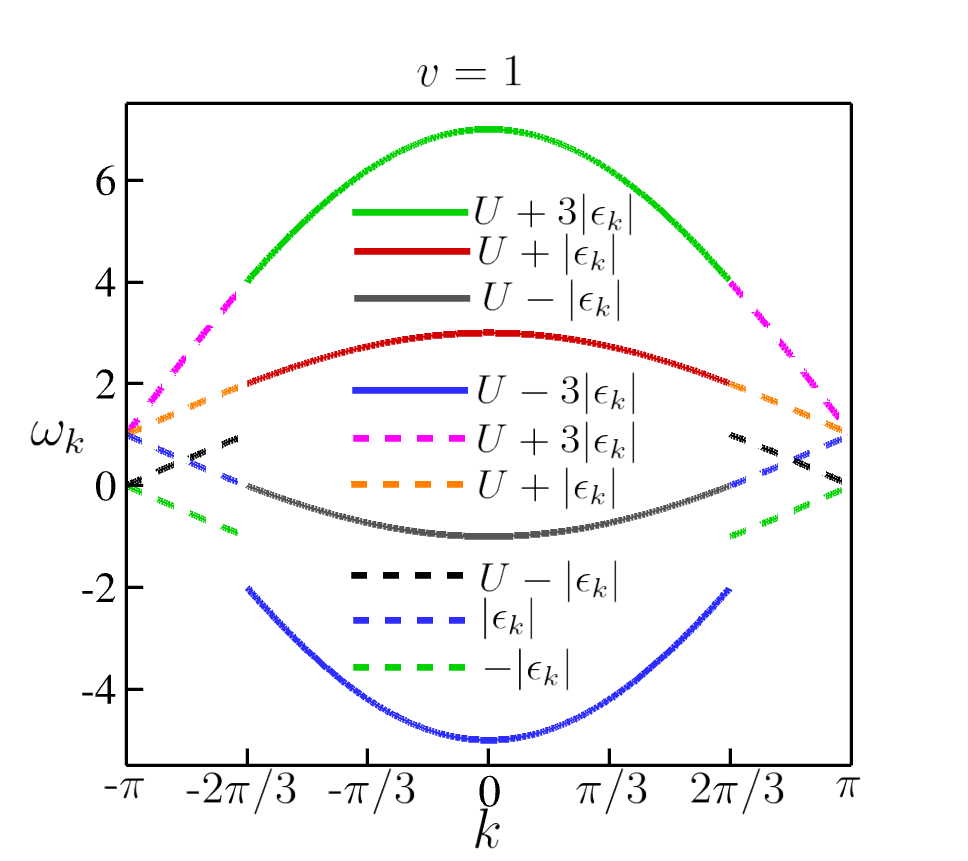}
		\includegraphics[scale=0.35]{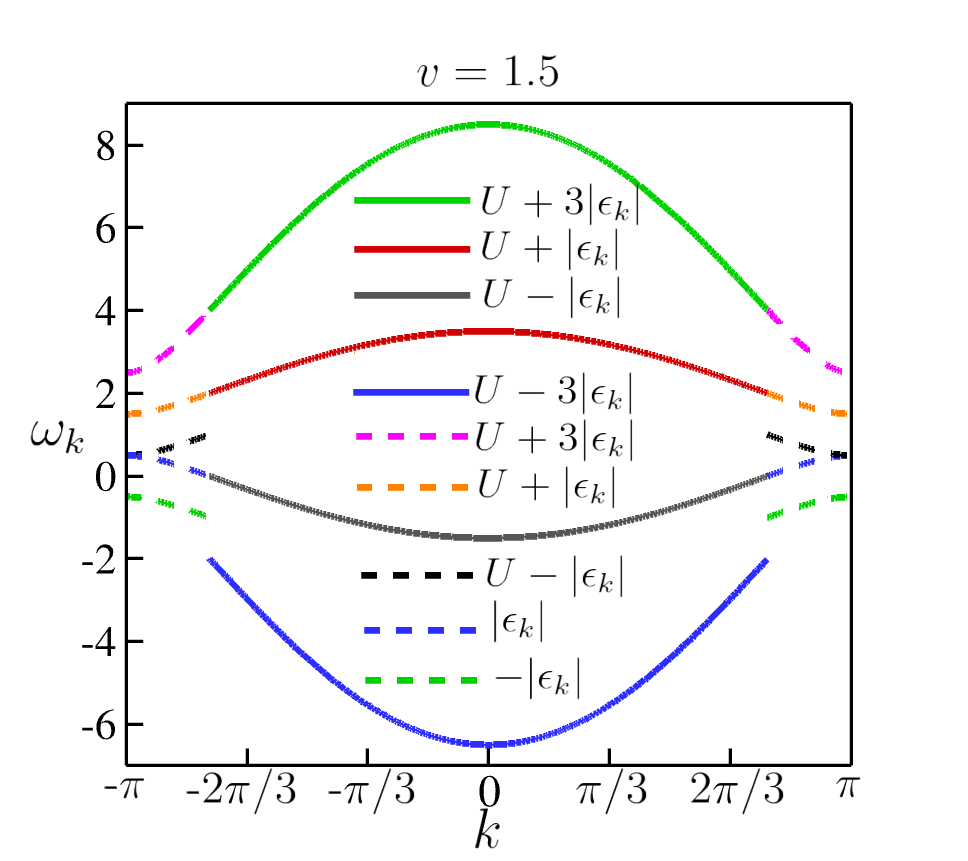}
		\caption{Energy of excitations in the sector with $1<n<2$. All plots are generated with the HK interaction set to $U=1$, while the intra-cell hopping $v$ varies from $0.5$ to $1.5$. Here, the inter-cell hopping $w$ is set to 1.} 
\label{Fig: energy-1n2}
\end{figure}
In the sector with $1<n<2$, a level-crossing occurs between the states corresponding to $n=1$ and $n=2$. In this case, the spectral function represents a combination of those from the 1-particle and 2-particle sectors. The energy of excitations in this region differs from those in the 1- and 2-particle sectors, particularly in the distribution of excitation bands with $\omega<0$. Specifically, around the center of the Brillouin zone, there are two bands with negative energy, while near the edge of the Brillouin zone, there is only one such band (see Fig. \ref{Fig: energy-1n2}). This contrasts with the 1- and 2-particle sectors, where the number of excitations with $\omega<0$ remains consistent at one and two, respectively, throughout the entire Brillouin zone.

The spectral weight, represents the probability of finding a particular excitation at a given energy. In the context of the HK model, the spectral weight provides important information about the distribution and strength of excitations within the system. Higher weights correspond to more probable excitations. In the HK model, the spectral weight can reveal how interactions between particles affect the distribution of energy levels. Changes in spectral weight can indicate phenomena such as the formation of quasiparticles or the presence of collective excitations. The spectral weight can be related to the density of states (DOS), showing how many states are available at a specific energy level. This helps in understanding the electronic structure and properties of the material.
The DOS of the excitation energy bands is expressed as:
\begin{align}\label{dos}
N(\omega)=4 \int \frac{dk}{2\pi} A(k,\omega),
\end{align}
where $4$ appeared above because $A^{aa}=A^{bb}=A^{ab}=A^{ba}$. As shown in Figs. \ref{Fig: DOSn1} and \ref{Fig: DOSn2}, the DOS derived from the spectral functions indicates the number of quasiparticles available at each excitation energy.

\section{Impact of chemical potential: Lifshitz transitions}\label{sec: Chemical}
\begin{figure}[!h]
	\centering
	\includegraphics[scale=0.18]{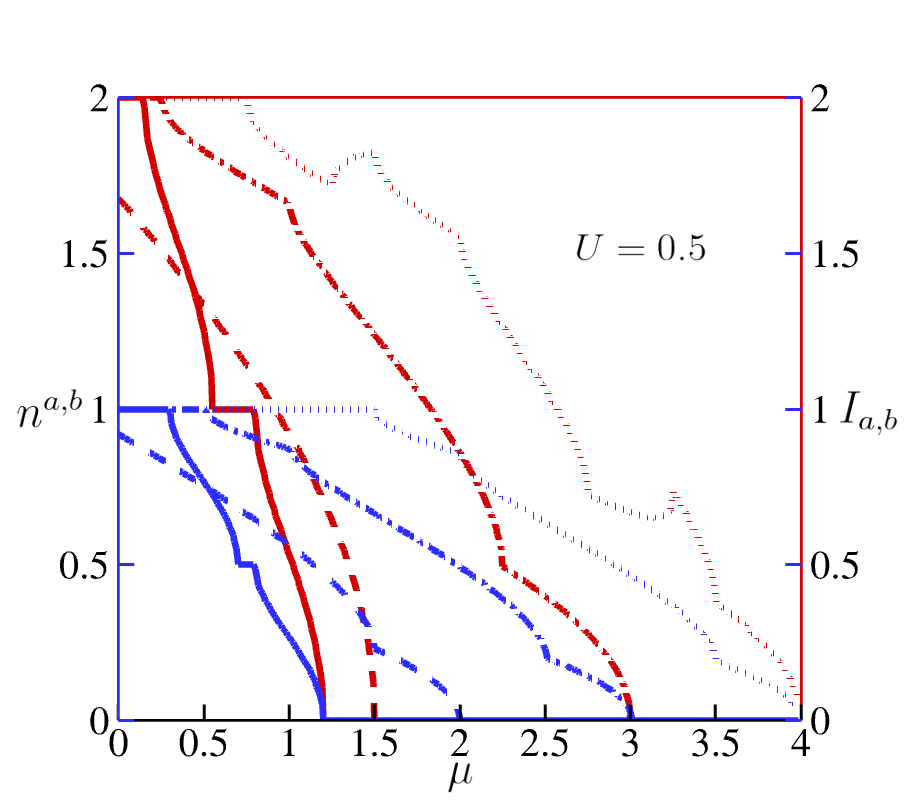}
	\includegraphics[scale=0.18]{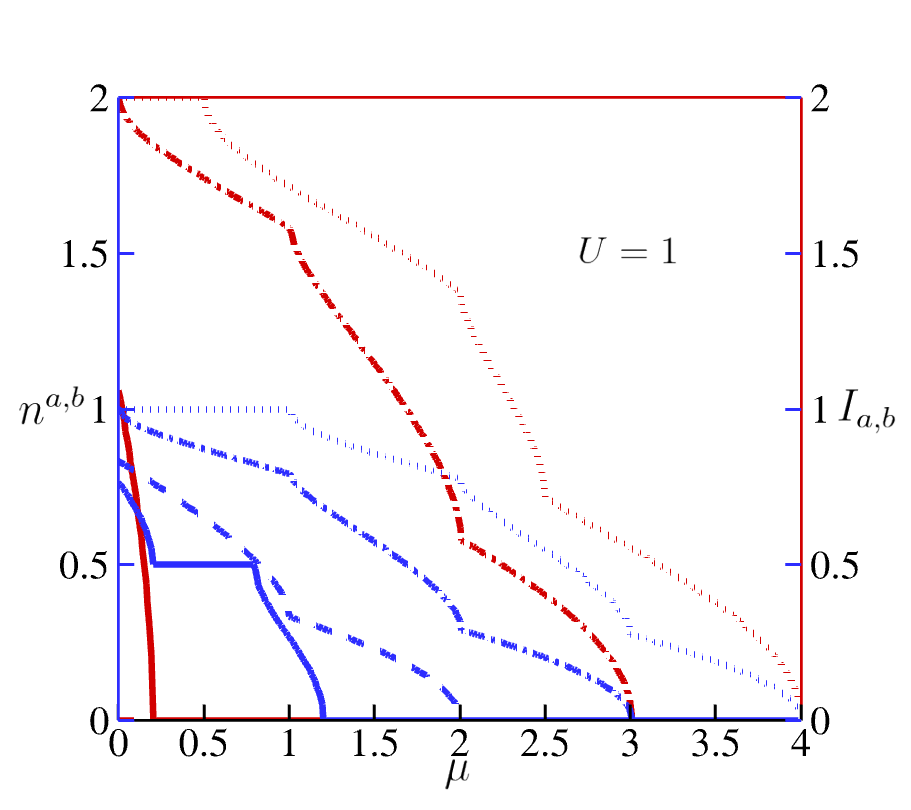}\\
	\includegraphics[scale=0.18]{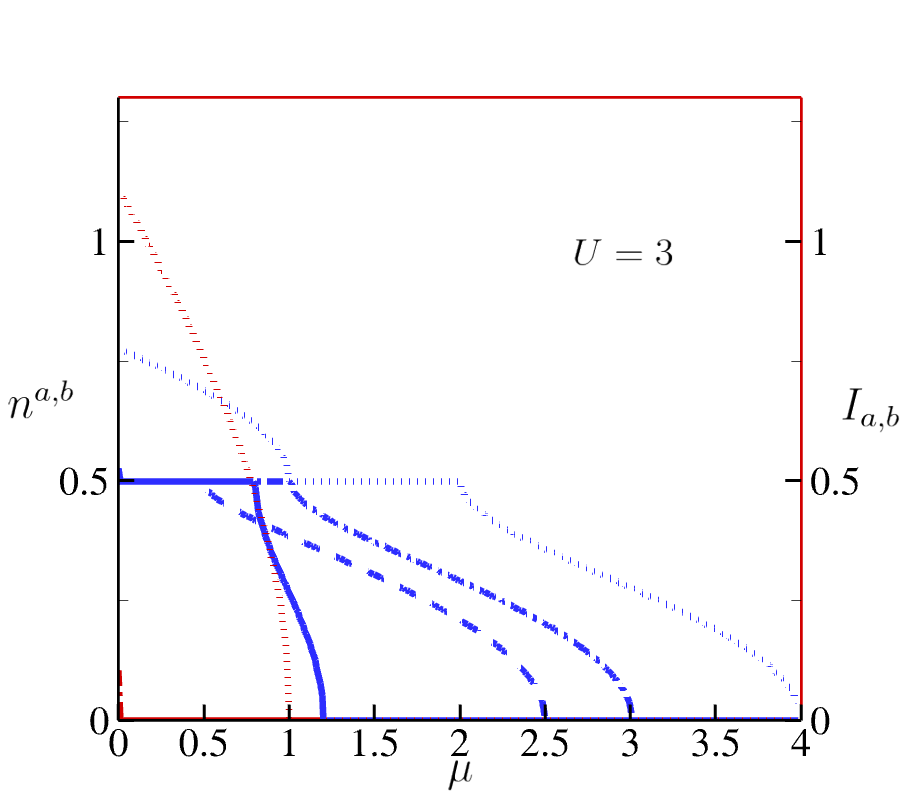}
	\caption{\small 
		The particle densities and Luttinger integrals as a function of $\mu$ for $w=1$.
		The solid lines are for $v=0.2$, dashed lines are for $v=1$, dashed-dot lines are for $v = 2.01$, and dotted lines are for $v=3$. Here, $n^a=n^b$, and $I_a=I_b$. At the coordinates $(U, v, \mu) = (0.5, 1, 1.44)$, $(U, v, \mu)=(1, 0.2, 0.11)$, and $(U, v, \mu)=(3, 3, 0.61)$ the ground state is a Fermi liquid.  	 
	} 
	\label{Fig:n-I-vs-mu}
\end{figure}
\begin{figure}[!h]
	\centering
	\includegraphics[scale=0.18]{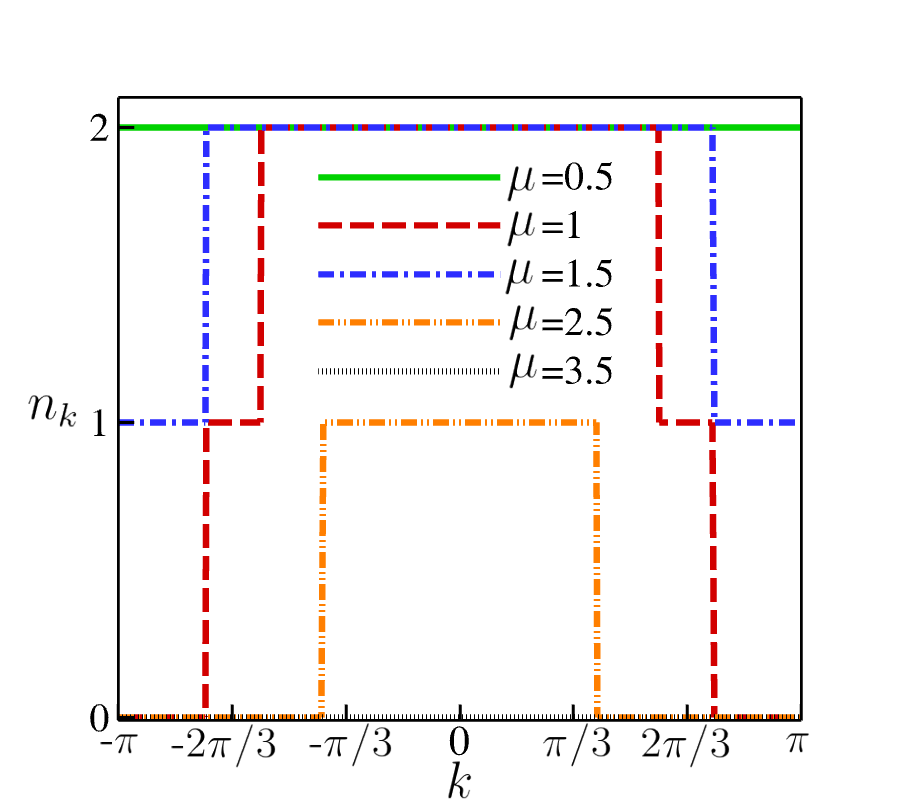}
	\caption{\small 
	Particle density ($n_k$) as a function of $k$ for various values of the chemical potential $\mu$ at $U=0.5$ and $v=2.01$.	} 
	\label{Fig:fermisurface-mu}
\end{figure}
In the presence of a chemial potential the terms $\mu \sum_{\sigma}(a_{k\sigma}^\dagger a_{k\sigma} + b_{k\sigma}^\dagger b_{k\sigma})$ are added to the Hamiltonian in Eq. \eqref{Eq:Hamilton-Kspace}, thus the diagonal terms of the Hamiltonian are modified. Therefore all the eigenvectors remain unchanged but the eigenvalues change. For instance, the eigenenergies of the Hamiltonain in 1-particle sector modifies to $\mu \pm|\epsilon_k|$, the eigenenergies of the Hamiltonian in 2-particle sector become $2\mu$, $2\mu+U$, and $2\mu+ U\pm2|\epsilon_k|$, the eigenenergies of the Hamiltonian in 3-particle sector become $3\mu +2U\pm|\epsilon_k|$, and in 4-particle sector the eigenenergy is $4\mu+3U$.

In the absence of the chemical potential, particle density varies between $1\leq n\leq 2$. 
Adding a chemical potential to the Hamiltonian can reduce the particle density below $n=1$. As illustrated in Fig. \ref{Fig:fermisurface-mu}, at certain values of $\mu$, such as $\mu=1.5$ (for $U=0.5$ and $v=2.01$), the behavior of $n_k$ as a function of $k$ exhibits a stepwise transition between the values of 0, 1, and 2. Specifically, near $k=0$, $n_k$ is consistently 2, transitioning to 1 in an intermediate region, and ultimately dropping to 0 in an outer region. These distinct regions are delineated by Fermi-like surfaces.

In Fig. \ref{Fig:n-I-vs-mu} we have plotted the particle densities $n^a$ and $n^b$ as a function of $\mu$, for various values of $v$ and $U$. Depending on the strengths of $v$ and $U$, the particle density can drop to zero at a certain value of $\mu$. 
In Fig. \ref{Fig:n-I-vs-mu}, we present the Luttinger integrals $I_a$ and $I_b$ as functions of $\mu$. It is evident that, akin to the scenario with $\mu=0$, the ground state remains a Fermi liquid only along a specific line. However, the position of this line in the $v-U$ space is altered by the presence of $\mu$.

The kink-like features in the particle density curve as a function of chemical potential $\mu$ suggest the occurrence of a Lifshitz transition, attributable to alterations in the Fermi or Fermi-like surface. This is further corroborated by the nonmonotonic behavior of the Luttinger integral (see Fig. \ref{Fig:n-I-vs-mu}), which indicates changes in the topology of the Fermi-like surface. At these critical points, the validity of the Luttinger theorem is compromised, resulting in a NFL ground state.

\begin{figure}[!h]
	\centering
    \includegraphics[scale=0.18]{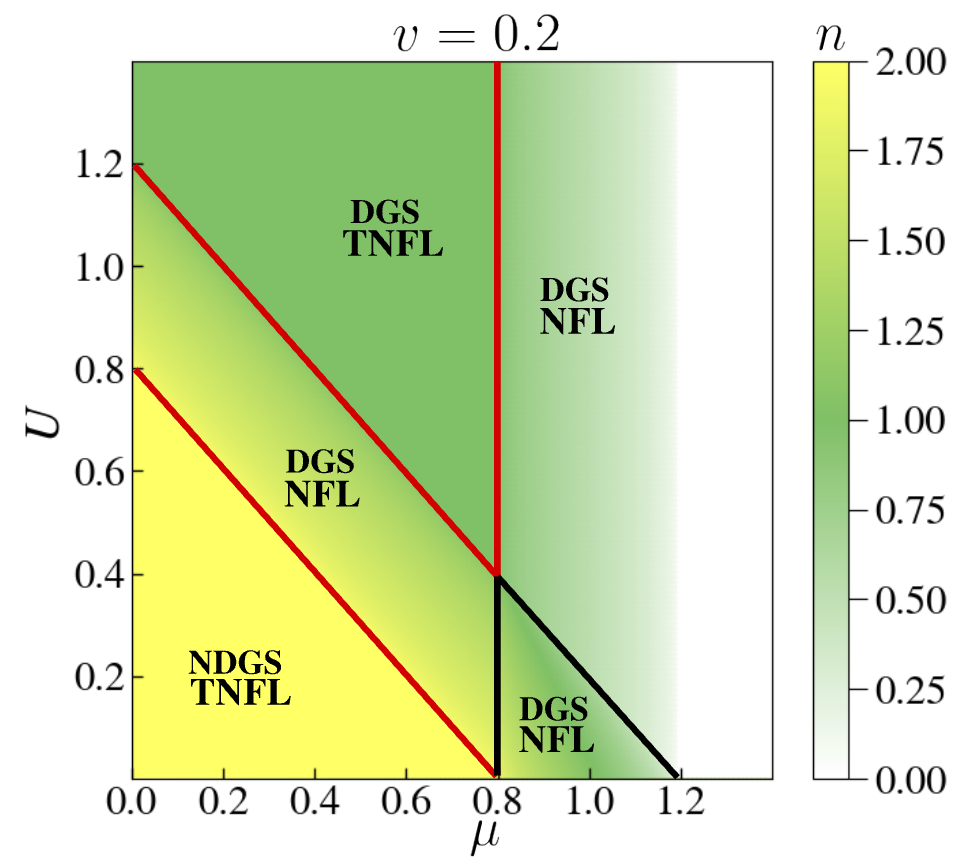}
	\includegraphics[scale=0.17]{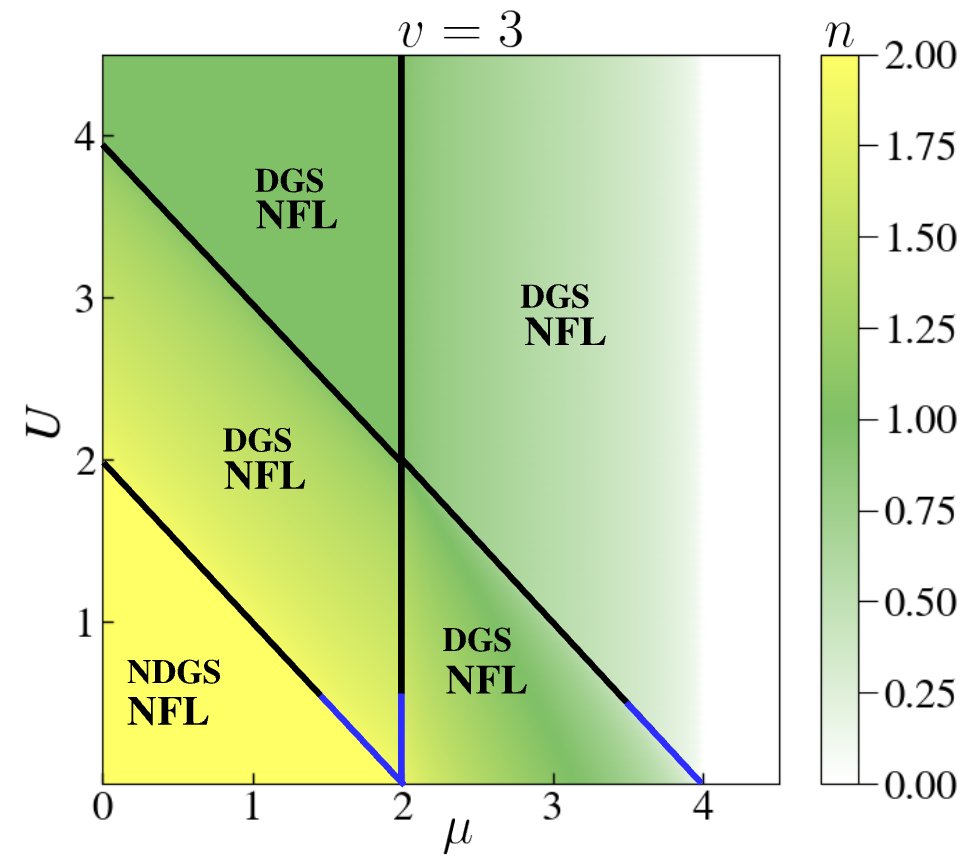}
	\caption{\small 
		The ground state phase diagram of the SSH-HK model in the presence of a chemical potential. Top: $v=0.2$, and bottom: $v=3$. Here, $w$ is set to 1. Along the black lines, continuous phase transitions from NDGS to DGS occur in the system. For $n=1$ and $n=2$, the system is topological when $v<1$ and trivial when $v>1$. The red line separates the topological NFL from the trivial one. For fractional $n$ and $n=0$, the system is trivial NFL. The ground state at the coordinates $(U=1, v=0.2, \mu=0.11)$ and $(U=3, v=3, \mu=0.61)$ is a Fermi liquid. These specific points are not depicted in the phase diagram. For $v=3$, Lifshitz transitions take place in the system at low interaction strengths (indicated by the blue lines).}
\label{phasewithmu}
\end{figure}

In Fig. \ref{phasewithmu}, we have illustrated the ground state phase diagram of the SSH-HK system for various values of intracell hopping $v$. 
The addition of the chemical potential $\mu$ to the diagonal elements of the Hamiltonian $H(k)$ does not influence the topological characteristics of the ground states corresponding to particle densities $n=1$ and $n=2$. Consequently, for $v<1$ the NFL phases with $n=1$ and $n=2$ retain their topological nature. In contrast, ground states associated with fractional or zero particle densities are classified as trivial. For $v>1$, the entirety of the ground state phase diagram transitions into a trivial regime.

\section{Conclusion}\label{sec: Conclude}

In this study, we rigorously investigated the phase diagram of the SSH chain under the influence of the HK interaction, characterized by long-range interactions in real space and momentum-space localization. This framework enabled us to perform exact diagonalization of the Hamiltonian, allowing for a comprehensive exploration of the ground state phase diagram of the SSH-HK model across various particle number sectors.

We categorized the system based on particle densities, identifying three distinct regions in the absence of a chemical potential: $n=1$, $1<n<2$, and $n=2$. Within each particle sector, we investigated the ground state properties utilizing the Luttinger integral, revealing that, with the exception of a specific line in the $1<n<2$ phase, the ground state predominantly exhibits NFL characteristics. Upon introducing a chemical potential, we expanded the phase diagram to include regions with $n=0$ and $0<n<1$. 
Significantly, the particle density as a function of the chemical potential $\mu$ exhibits pronounced kink-like features, indicative of a Lifshitz transition, which arises from alterations in the topology of the Fermi surface.

In the absence of interactions, the SSH chain demonstrates a topological phase when the intracell hopping parameter $v$ is smaller than the intercell hopping parameter $w$. The introduction of the HK interaction modifies this scenario significantly. We defined a many-body Zak phase to investigate the topological properties of the SSH-HK model across different particle number sectors. In the one-particle sector, when $v<w$, the ground state is identified as a topological NFL characterized by a many-body Zak phase of $2\pi$. Beyond the critical point $v>w$, the topological NFL transitions into a trivial insulator, reflecting the behavior of the SSH chain. In the two-particle sector with $U=0$, the point $v=w$ serves as a critical point between the topological NFL and the trivial NFL, with the HK interaction ($U\neq 0$) creating a boundary that connects the topological NFL with the NFL in the $1<n<2$ sector, forming a diagonal line in the phase diagram.

We also established that, similar to the topological phase of the SSH model where the Zak phase is proportional to electronic polarization, the many-body Zak phase, derived from the ground state in momentum space, serves as a direct measure of the electronic polarization in the SSH-HK model's topological NFL phase. 

To further elucidate the excitations in the NFL phases observed in the ground state phase diagram of the SSH-HK model, we analyzed the spectral functions and density of states. In both the 1- and 2-particle sectors, the spectral functions exhibited multiple delta functions with varying spectral weights. The 1-particle sector revealed five distinct quasiparticle excitations, including one with negative energy and four with positive energies. In contrast, the 2-particle sector displayed four well-defined quasiparticles, comprising two hole-like (negative energy) and two electron-like (positive energy) states. In the $1<n<2$ sector, the ground state is constructed from the ground states of two sectors, with negative energy excitations partitioned into segments occupied by one or two particles, delineated by a quasi-Fermi surface.

Our findings significantly enhance the understanding of topological NFLs and their potential applications in high-temperature superconductivity and other correlated electron systems. The intricate interplay between topology and strong electron correlations in these systems may offer new pathways for realizing and harnessing exotic phases of matter. The insights gained from this work may pave the way for future explorations into the role of topological features in strongly correlated systems, potentially leading to the discovery of novel quantum materials and phenomena.


\bibliographystyle{ieeetr}
\bibliography{scibib}

\end{document}